\documentclass[]{interact}

\usepackage[numbers,sort&compress]{natbib}
\bibpunct[, ]{[}{]}{,}{n}{,}{,}

\makeatletter
\def\NAT@def@citea{\def\@citea{\NAT@separator}}
\makeatother
\usepackage{epstopdf}
\usepackage[caption=false]{subfig}
\theoremstyle{plain}

\theoremstyle{definition}

\theoremstyle{remark}

 \usepackage{amsmath,amssymb,bm,euscript}
 \usepackage{graphicx,boxedminipage}
 \usepackage[bf,small]{caption2}
 \usepackage{floatflt}
 \usepackage[hypertex]{hyperref}
 \usepackage{xcolor,fancybox}
\definecolor{Dark-blue}{RGB}{0,0,255}
\newcommand{\bra}[1]{\ensuremath{\bm{\langle}#1\bm{|}}}
\newcommand{\ket}[1]{\ensuremath{\bm{|}#1\bm{\rangle}}}
\newcommand{\Tr}{\ensuremath\mathrm{Tr\,}}

\begin{document}

\title{A method of effective potentials for calculating the frequency spectrum of eccentrically layered spherical cavity resonators}

 \author{
 \name{Z.~E. Eremenko, Yu.~V. Tarasov\thanks{CONTACT Yu.~V. Tarasov. Email: yuriy.tarasov@gmail.com}, and I.~N. Volovichev}
 \affil{A.~Ya. Usikov Institute for Radiophysics and Electronics, NAS of Ukraine,\\
  12 Proskura Str., 61085 Kharkov, Ukraine}
 }

\maketitle

%------------------------------------------------------------------
\begin{abstract}
  A novel method for the calculation of eigenfrequencies of non-uniformly filled spherical cavity resonators is developed. The impact of the system symmetry on the electromagnetic field distribution as well as on its degrees of freedom (the set of resonant modes) is examined. It is shown that in the case of angularly symmetric cavity, regardless of its radial non-uniformity, the set of resonator modes is, as anticipated, a superposition of TE and TM oscillations which can be described in terms of a single scalar function independently of each other. The spectrum is basically determined through the introduction of effective ``dynamic'' potentials which encode the infill inhomogeneity. The violation of polar symmetry in the infill dielectric properties, the azimuthal symmetry being simultaneously preserved, suppresses all azimuthally non-uniform modes of electric-type (TM) oscillations. In the absence of angular symmetry of both electric and magnetic properties of the resonator infill, only azimuthally uniform distribution of both TM and TE fields is expected to occur in the resonator. The comparison is made of the results obtained through the proposed method and of the test problem solution obtained with use of commercial solvers. The method appears to be efficient for computational complex algorithms for solving spectral problems, including those for studying the chaotic properties of electrodynamic systems' spectra.
\end{abstract}
%-------------------------------------------------------------------

% \allowdisplaybreaks
%=======================================
\section{Introduction}
\label{Intro}
%=======================================

Isolated wave systems of macroscopic and mesoscopic dimensions, in particular, quasi-optical electromagnetic (EM) resonators arouse large interest of many researchers due to their numerous applications and fundamental physical properties. This relates, specifically, to metallic cavities filled with an anisotropic or gyroelectric medium \cite{bib:Zouros17a,bib:Zouros17b,bib:KolezasZouros18}, metamaterial cavities \cite{bib:BozzaOlivRaff07}, oscillators and filters \cite{bib:JuliGuil86}, the measurement of the ferrite materials' dielectric tensor \cite{bib:Dankov06,bib:OkadaTanaka91}, etc. EM cavity resonators are also used quite extensively to simulate quantum mechanical systems of finite dimension, both open and closed. The main tool to examine their chaotic properties, just the same as for classical dynamic systems, is the statistical analysis of their spectra \cite{bib:Stockman99}. Most of the conclusions of the statistical theory regarding the dynamical system spectra settle on their symmetry properties \cite{bib:BohigGian75,bib:BohigGianSchmit84}, since direct calculation of the spectral parameters is normally difficult to perform. Similar methods apply to wave systems as well, and the conclusions about their chaotic properties are mostly made on the basis of ray (``trajectory'') optics \cite{bib:BarYanNaydKur06}. Such an approach, in view of significant uncertainty in the position of indicative points on the ray trajectories, if taking account of finiteness of the wave length, cannot provide sufficient reliability of the results, especially when it comes to closed systems such as cavity resonators. Besides, to obtain the reliable spectra of resonators with some inhomogeneities in the bulk it appears to be, as a rule, computationally expensive task, if one considers the number of (Maxwell's) equations subject to the solution.

The latter circumstance has previously been overcome for homogeneous systems through their examination in terms of Hertz functions \cite{bib:Stratton07} and/or Debye potentials \cite{bib:Vainstein88} rather than directly through EM field components. The opportunity to express all vector-field components through two scalar functions was noted back in a long time publication by  Whittaker \cite{bib:Whittaker03} (more detailed reference list can be found in paper by Nisbet \cite{bib:Nisbet55}). The approach has proven to be quite productive, and not for uniform systems only but also for some systems with gyrotropic properties~\cite{bib:Przezdziecki_1,bib:Przezdziecki_2,bib:Weiglhofer2000}. Yet, for a long time it was not possible to extend the approach to arbitrary nonhomogeneous systems, which has substantially reduced the capability of the method. One of the few successful attempts to overcome this issue was made in Ref.~\cite{bib:Malykh_etal_17}, where the method for solving the Helmholtz equation in heterogeneous closed waveguides was developed by reducing it to the Hamiltonian form through the introduction of \emph{four} scalar potentials. In our study we propose a novel method to calculate the spectra of inhomo\-geneous cavity resonators by introducing \emph{a couple} of Debye-type scalar \emph{effective potentials}, with the subsequent solution of the emergent perturbed Helmholtz equations within the whole resonator rather than in its partial regions. The potentials originate from two Hertz vectors through which all the EM field components are expressed in the system of any geometry \cite{bib:Stratton07}. The scalar nature of these potentials relates to the fact that in the spherical and some other coordinate systems one component only of each of the two Hertz vectors suffices to represent through them the entire set of vector field components \cite{bib:Vainstein88}.

To solve the Helmholtz equations subject to the perturbation potentials we apply the method of oscillation mode separation applicable for both homo- and heterogeneous systems \cite{bib:GanErTar07}, which is a strict procedure, in contrast to the approximate separation normally performed within the framework of the conventional cou\-pled-mode theory~\cite{bib:HausHuang91}. The method we suggest for mode decomposition involves the solution of matrix equations either to directly find their eigenvalues or to calculate the Green function containing the full information regarding the system spectrum. In both cases, it is far from easy to find the practical analytical solution, so at some stage it is necessary to apply also some numerical methods. For this, in its turn, it is necessary to deliberate some crucial issues. In the theory set forth below, the expansion of the Debye-type scalar potentials is carried out over an infinite orthonormal sets of basis functions. Obviously, in computer calculations one has to limit oneself to a finite number of those functions. To do this, it is necessary, for one thing, to determine the sensitivity of the spectrum to the dimension of the reduced basis set. Secondly, there appears a need to analyze the accuracy and the stability of the computational scheme when finding the matrix elements of the effective potentials. And finally, the calculation of the spectrum in a sufficiently large frequency interval implies the consumption of quite a~lot of processor time. So, it is important to choose the efficient computational algorithm, in particular, for numerical integration in 3D. In our paper, the speculations and the results on this subjects are presented as well.

In the present paper we expound the general method for calculating the spectra of spherical cavity resonators and the results that relate to the specific kind of inhomogeneity, namely, the inhomogeneity in the form of concentric sharp-bounded layers. We also discuss the effect of the resonator infill symmetry on the number of degrees of freedom of the EM field, and the possibility to represent the resonator eigenmodes in the form of TE and TM components.

%=======================================
\section{Statement of the problem and derivation of basic equations}
\label{StateProb}
%=======================================

\begin{figure}[h]
 \captionstyle{flushleft}
 \centering
 \scalebox{.5}[.5]{\includegraphics{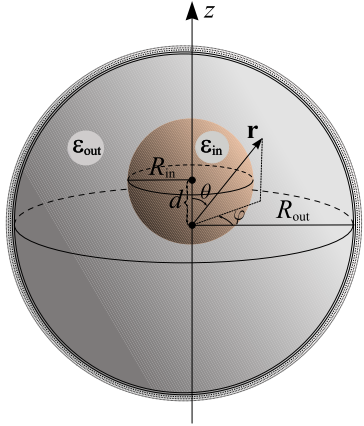}}
 \caption{The eccentrically layered spherical cavity resonator.}
 \label{DoubleSpherRes}
\end{figure}

The primary goal of our study is to determine the degree of chaos in the spectrum of a spherical EM cavity resonator of radius $R_{\mathrm{out}}$ filled with a piecewise homogeneous dielectric substance, as shown in Figure~\ref{DoubleSpherRes}. The permittivity of the outer dielectric layer is $\varepsilon_{\mathrm{out}}$
whereas the permittivity of the inclusion, which we assume to be also the dielectric sphere yet of radius $R_{\mathrm{in}}$ and the center shifted to arbitrary distance $d$ from the center of outer sphere, is $\varepsilon_{\mathrm{in}}$. The permeability of both parts of the resonator infill is assumed to be the same, and we put it equal to unity everywhere in the final formulae.

Normally, to find the EM fields in a system with heterogeneously distributed electric and magnetic parameters one has to solve the set of coupled Maxwell equations, which in the stationary case is represented by the following \textit{eight} scalar equalities,
\begin{subequations}\label{Maxwell_EH}
\begin{align}
\label{rotE}
 & \mathrm{rot}\,\mathbf{E}=ik\,\mu(\mathbf{r})\mathbf{H}\ ,\\
\label{rotH}
 & \mathrm{rot}\,\mathbf{H}=-ik\,\varepsilon(\mathbf{r})\mathbf{E}\ ,\\
\label{divD}
 & \mathrm{div}\big[\varepsilon(\mathbf{r})\mathbf{E}\big]=0\ ,\\
\label{divB}
 & \mathrm{div}\big[\mu(\mathbf{r})\mathbf{H}\big]=0\ .
\end{align}
\end{subequations}
Here, $k=\omega/c$, $\varepsilon(\mathbf{r})$ and $\mu(\mathbf{r})$ are the permittivity and the permeability of the resonator infill, both of which at this stage are assumed to be quite arbitrary functions of the coordinate vector~$\mathbf{r}$. The number of equations to be solved may be reduced if one goes over from the vector-valued electric and magnetic fields to the scalar-valued Hertz functions. Until now, these functions have been consistently introduced for entirely homogeneous systems \cite{bib:Stratton07,bib:Vainstein88} and for gyrotropic systems with distinguished axis \cite{bib:Przezdziecki_1,bib:Przezdziecki_2,bib:Weiglhofer2000}. In our study we show that the analogous Hertzian functions can be introduced for arbitrarily heterogeneous systems as well, and without any special restrictions on the degree and the character of inhomogeneity.

Following Refs.~\cite{bib:Stratton07,bib:Vainstein88}, we represent the fields $\mathbf{E}$ and $\mathbf{H}$ as the sums of the components of ``electric'' (\emph{e}) and ``magnetic'' (\emph{m}) type,
\begin{subequations}\label{EH=electr+magn}
\begin{align}
\label{E=electr+magn}
  \mathbf{E} & =\mathbf{E}^e+\mathbf{E}^m\ ,\\
\label{H=electr+magn}
  \mathbf{H} & =\mathbf{H}^e+\mathbf{H}^m\ ,
\end{align}
\end{subequations}
both satisfying the set of equations \eqref{Maxwell_EH} independently. These components, in view of the particular symmetry of the set \eqref{Maxwell_EH}, may, in principle, be defined through different, yet symmetrical, dependencies on the conventional vector potential $\mathbf{A}(\mathbf{r})$ and the scalar potential $\varphi(\mathbf{r})$. In our study we prefer a~different definition, namely, trough the vanishing of the \textit{radial} components of magnetic field $\mathbf{H}^e$ and electric field $\mathbf{E}^m$. In accordance with this definition, the fields marked by indexes \emph{e} and \emph{m} will be termed the TM and TE polarized fields, respectively. In the uniform and unbounded medium these fields are uncoupled by definition. In the cavity resonator, however, the boundary conditions (BCs) couple them, so it may seem not to make much sense to distinguish the fields by their polarization. Yet, below it will be shown that such a distinction is quite reasonable since there is the possibility to satisfy the BCs for TM and TE polarized fields independently of each other.

The practicality of the above definition comes to be appreciable if one writes down  equations~\eqref{Maxwell_EH} in spherical coordinates associated with the center of outer sphere, see Figure~\ref{DoubleSpherRes}. In this curvilinear system equations \eqref{rotE} and \eqref{rotH} take the form
\begin{subequations}\label{MaxEqs-sphere}
\begin{align}
\label{MaxEq-Er}
 & \frac{\partial E_r}{\partial\varphi}-
 \frac{\partial}{\partial r}\big(r\sin\vartheta\, E_{\varphi}\big)=
 ik\,\mu(\mathbf{r})\,r\sin\vartheta H_{\vartheta}\ ,\\
\label{MaxEq-Etheta}
 & \frac{\partial}{\partial r}\big(r E_{\vartheta}\big)-
 \frac{\partial E_r}{\partial\vartheta}=
 ik\,\mu(\mathbf{r})\,r H_{\varphi}\ ,\\
\label{MaxEq-Ephi}
 & \frac{\partial}{\partial\vartheta}\big(r\sin\vartheta\, E_{\varphi}\big)-
 \frac{\partial}{\partial\varphi}\big(r E_{\vartheta}\big)=
 ik\,\mu(\mathbf{r})\,r^2\sin\vartheta\, H_r\ ,\\
\label{MaxEq-Hr}
 & \frac{\partial H_r}{\partial\varphi}-
 \frac{\partial}{\partial r}\big(r\sin\vartheta\, H_{\varphi}\big)=
 -ik\,\varepsilon(\mathbf{r})\,r\sin\vartheta E_{\vartheta}\ ,\\
\label{MaxEq-Htheta}
 & \frac{\partial}{\partial r}\big(r H_{\vartheta}\big)-
 \frac{\partial H_r}{\partial\vartheta}=
 -ik\,\varepsilon(\mathbf{r})\,r E_{\varphi}\ ,\\
\label{MaxEq-Hphi}
 & \frac{\partial}{\partial\vartheta}\big(r\sin\vartheta\, H_{\varphi}\big)-
 \frac{\partial}{\partial\varphi}\big(r H_{\vartheta}\big)=
 -ik\,\varepsilon(\mathbf{r})\,r^2\sin\vartheta\, E_r\ .
\end{align}
\end{subequations}
It looks quite appreciable to consider this set of equations separately by choosing either $H_r\equiv 0$ or $E_r\equiv 0$. In such a~way we would split a general solution into the sum of solutions of two types which are conventionally designated as the TM and TE polarized fields for a spherical resonator~\cite{bib:Vainstein88}.

Instead of being written in terms of double six EM field components marked by indices ($e$) and ($m$), the set of equations~\eqref{MaxEqs-sphere} may be rewritten in terms of two scalar fields  only, $U(\mathbf{r})$ and $V(\mathbf{r})$, which are referred to as the electric and magnetic Hertz functions, respectively. For \emph{e}-type fields in Eqs.~\eqref{EH=electr+magn} we will use the substitution formulas
\begin{subequations}\label{Substitutions}
\begin{equation}\label{Ephi_Etheta_Er->U}
 \begin{aligned}
 & E_{\varphi}^e=\frac{1}{\varepsilon(\mathbf{r})\,r\sin\vartheta}\cdot\frac{\partial^2 U} {\partial\varphi\partial r}\ , \qquad
  E_{\vartheta}^e=\frac{1}{\varepsilon(\mathbf{r})\,r}\cdot
  \frac{\partial^2U}{\partial\vartheta\partial r}\ ,\\[6pt]
 &  E_r^e=-\frac{1}{\varepsilon(\mathbf{r})\,r^2\sin\vartheta}
  \left[\frac{\partial}{\partial\vartheta}\left(\sin\vartheta\frac{\partial U} {\partial\vartheta}\right)+ \frac{1}{\sin\vartheta}\frac{\partial^2U}{\partial\varphi^2}\right]\ ,\\[6pt]
 & %\hspace{25pt}
  H_{\varphi}^e=\frac{ik}{r}\frac{\partial U}{\partial\vartheta}\ ,\qquad
  H_{\vartheta}^e=\frac{-ik}{r\sin\vartheta}\frac{\partial U}{\partial\varphi}\ ,\qquad
  H_r^e\equiv 0\ ,
 \end{aligned}
\end{equation}
which define the transverse-magnetic (TM) part of the entire EM field, while for the field components of \emph{m}-type (TE field) the substitutions will be applied
\begin{equation}\label{Hphi_Htheta_Hr->V}
 \begin{aligned}
 & H_{\varphi}^m=\frac{1}{\mu(\mathbf{r})\,r\sin\vartheta}\cdot\frac{\partial^2 V} {\partial\varphi\partial r}\ , \qquad
  H_{\vartheta}^m=\frac{1}{\mu(\mathbf{r})\,r}\cdot
  \frac{\partial^2V}{\partial\vartheta\partial r}\ ,\\[6pt]
 &  H_r^m=-\frac{1}{\mu(\mathbf{r})\,r^2\sin\vartheta}
  \left[\frac{\partial}{\partial\vartheta}\left(\sin\vartheta\frac{\partial V} {\partial\vartheta}\right)+ \frac{1}{\sin\vartheta}\frac{\partial^2V}{\partial\varphi^2}\right]\ ,\\[6pt]
 & %\hspace{25pt}
  E_{\varphi}^m=-\frac{ik}{r}\frac{\partial V}{\partial\vartheta}\ ,\qquad
  E_{\vartheta}^m=\frac{ik}{r\sin\vartheta}\frac{\partial V}{\partial\varphi}\ ,\qquad
  E_r^m\equiv 0\ .
 \end{aligned}
\end{equation}
\end{subequations}

The motivation for both of these kinds of field representation is mostly taken from Ref.~\cite{bib:Vainstein88}, where in the case of a uniform medium, with $\varepsilon=\mathrm{const}$ and $\mu=\mathrm{const}$, substitution of \eqref{Ephi_Etheta_Er->U} and \eqref{Hphi_Htheta_Hr->V} into Maxwell equations \eqref{MaxEqs-sphere} has resulted in the identical Helmholtz equations for Debye potentials (DPs) $u(\mathbf{r})=U(\mathbf{r})/r$ and $v(\mathbf{r})=V(\mathbf{r})/r$, namely,
\begin{equation}\label{U,V-Helm}
  \Big(\Delta+k^2\varepsilon\mu\Big)
  \begin{pmatrix}
   u\\v
  \end{pmatrix}
  =0\ .
\end{equation}
The nonuniformity of the medium, if any, may be, in principle, included by putting into the right-hand-sides of equations~\eqref{U,V-Helm} the ``source'' terms in the form of either electric or magnetic polarization \cite{bib:Nisbet55}. In the present study, however, we suggest a different way for considering the medium heterogeneity, namely, by merely replacing the constant $\varepsilon$ and $\mu$ with variable functions $\varepsilon(\mathbf{r})$ and $\mu(\mathbf{r})$, as shown in Eqs.~\eqref{Ephi_Etheta_Er->U}. Certainly, we thus arrive at equations for $U$ and $V$ that differ from simple equations~\eqref{U,V-Helm} by additional terms accounting for medium heterogeneity. Yet, regarding these terms as some kind of ``dynamic'' potentials we can take them into account as a perturbation for the standard Helmholtz operator.

With substitutions \eqref{Ephi_Etheta_Er->U}, Maxwell equations \eqref{MaxEq-Er} and \eqref{MaxEq-Etheta} acquire the form
\begin{subequations}\label{Two_eqs_for_U-full}
\begin{align}\label{1st_eq_for_U-full}
&  \frac{\partial}{\partial r}\left[\frac{1}{\varepsilon(\mathbf{r})}
  \frac{\partial^2U}{\partial\varphi\partial r}\right]+
  \frac{1}{r^2}\frac{\partial}{\partial\varphi}
  \left\{\frac{1}{\varepsilon(\mathbf{r})}
  \left[\frac{1}{\sin\vartheta}\frac{\partial}{\partial\vartheta}
  \left(\sin\vartheta\frac{\partial}{\partial\vartheta}\right) +
  \frac{1}{\sin^2\vartheta}\frac{\partial^2}{\partial\varphi^2}\right]U\right\}+
\notag\\
 & \phantom{\frac{\partial}{\partial r}\left[\frac{1}{\varepsilon(\mathbf{r})}
  \frac{\partial^2U}{\partial\varphi\partial r}\right]+
  \frac{1}{r^2}\frac{\partial}{\partial\varphi}
  \bigg\{\frac{1}{\varepsilon(\mathbf{r})}
  \bigg[\frac{1}{\sin\vartheta}\frac{\partial}{\partial\vartheta}
  \left(\sin\vartheta\frac{\partial}{\partial\vartheta}\right)
  \frac{\partial^2}{\partial\varphi^2} \quad }
  + k^2\mu(\mathbf{r})\frac{\partial U}{\partial\varphi}=0\ ,
 \\[6pt] % \intertext{and}
\label{2nd_eq_for_U-full}
 & \frac{\partial}{\partial r}\left[\frac{1}{\varepsilon(\mathbf{r})}
  \frac{\partial^2U}{\partial\vartheta\partial r}\right]+
  \frac{1}{r^2}\frac{\partial}{\partial\vartheta}
  \left\{\frac{1}{\varepsilon(\mathbf{r})}
  \left[\frac{1}{\sin\vartheta}\frac{\partial}{\partial\vartheta}
  \left(\sin\vartheta\frac{\partial}{\partial\vartheta}\right)+
  \frac{1}{\sin^2\vartheta}\frac{\partial^2}{\partial\varphi^2}\right]U\right\}+
\notag\\
 & \phantom{\frac{\partial}{\partial r}\left[\frac{1}{\varepsilon(\mathbf{r})}
  \frac{\partial^2U}{\partial\varphi\partial r}\right]+
  \frac{1}{r^2}\frac{\partial}{\partial\varphi}
  \bigg\{\frac{1}{\varepsilon(\mathbf{r})}
  \bigg[\frac{1}{\sin\vartheta}\frac{\partial}{\partial\vartheta}
  \left(\sin\vartheta\frac{\partial}{\partial\vartheta}\right)
  \frac{\partial^2}{\partial\varphi^2} \quad }
  + k^2\mu(\mathbf{r})\frac{\partial U}{\partial\vartheta}=0 ,
\end{align}
while equations \eqref{MaxEq-Hr} and \eqref{MaxEq-Htheta} are fulfilled identically. Equation \eqref{MaxEq-Hphi} also is fulfilled identically while Eq.~\eqref{MaxEq-Ephi} reduces to
\begin{align}\label{U-third_eq}
  \frac{\partial\varepsilon(\mathbf{r})}{\partial\vartheta} \cdot
  \frac{\partial^2 U}{\partial\varphi\partial r} -
  \frac{\partial\varepsilon(\mathbf{r})}{\partial\varphi}\cdot
  \frac{\partial^2 U}{\partial\vartheta\partial r}=0\ .
\end{align}
\end{subequations}
Analogously, substitution of \eqref{Hphi_Htheta_Hr->V} result in the following set of equations for the fields we recognize as the magnetic-type ones, namely,
\begin{subequations}\label{Two_eqs_for_V-full}
\begin{align}
\label{1st_eq_for_V-full}
 & \frac{\partial}{\partial r}\left[\frac{1}{\mu(\mathbf{r})}
  \frac{\partial^2V}{\partial\varphi\partial r}\right]+
  \frac{1}{r^2}\frac{\partial}{\partial\varphi}
  \left\{\frac{1}{\mu(\mathbf{r})}
  \left[\frac{1}{\sin\vartheta}\frac{\partial}{\partial\vartheta}
  \left(\sin\vartheta\frac{\partial}{\partial\vartheta}\right)+
  \frac{1}{\sin^2\vartheta}\frac{\partial^2}{\partial\varphi^2}\right]V\right\}+
\notag\\
 & \phantom{\frac{\partial}{\partial r}\left[\frac{1}{\varepsilon(\mathbf{r})}
  \frac{\partial^2U}{\partial\varphi\partial r}\right]+
  \frac{1}{r^2}\frac{\partial}{\partial\varphi}
  \bigg\{\frac{1}{\varepsilon(\mathbf{r})}
  \bigg[\frac{1}{\sin\vartheta}\frac{\partial}{\partial\vartheta}
  \left(\sin\vartheta\frac{\partial}{\partial\vartheta}\right)
  \frac{\partial^2}{\partial\varphi^2} \quad }
  +k^2\varepsilon(\mathbf{r})\frac{\partial V}{\partial\varphi}=0\ ,\\[6pt]
\label{2nd_eq_for_V-full}
 & \frac{\partial}{\partial r}\left[\frac{1}{\mu(\mathbf{r})}
  \frac{\partial^2V}{\partial\vartheta\partial r}\right]+
  \frac{1}{r^2}\frac{\partial}{\partial\vartheta}
  \left\{\frac{1}{\mu(\mathbf{r})}
  \left[\frac{1}{\sin\vartheta}\frac{\partial}{\partial\vartheta}
  \left(\sin\vartheta\frac{\partial}{\partial\vartheta}\right)+
  \frac{1}{\sin^2\vartheta}\frac{\partial^2}{\partial\varphi^2}\right]V\right\}+
\notag\\
 & \phantom{\frac{\partial}{\partial r}\left[\frac{1}{\varepsilon(\mathbf{r})}
  \frac{\partial^2U}{\partial\varphi\partial r}\right]+
  \frac{1}{r^2}\frac{\partial}{\partial\varphi}
  \bigg\{\frac{1}{\varepsilon(\mathbf{r})}
  \bigg[\frac{1}{\sin\vartheta}\frac{\partial}{\partial\vartheta}
  \left(\sin\vartheta\frac{\partial}{\partial\vartheta}\right)
  \frac{\partial^2}{\partial\varphi^2} \quad }
  +k^2\varepsilon(\mathbf{r})\frac{\partial V}{\partial\vartheta}=0\ ,\\[6pt]
\label{V-third_eq}
 & \phantom{\frac{\partial}{\partial r}\left[\frac{1}{\mu(\mathbf{r})}
  \frac{\partial^2 V}{\partial\vartheta\partial r}\right]+
  \frac{1}{r^2}\frac{\partial}{\partial\vartheta}\quad }
  \frac{\partial\mu(\mathbf{r})}{\partial\vartheta} \cdot
  \frac{\partial^2 V}{\partial\varphi\partial r}-
  \frac{\partial\mu(\mathbf{r})}{\partial\varphi} \cdot
  \frac{\partial^2 V}{\partial\vartheta\partial r}=0\ .
\end{align}
\end{subequations}

Besides curl-based equations \eqref{rotE} and \eqref{rotH}, which are now written in the form of Eqs.~\eqref{Two_eqs_for_U-full} and \eqref{Two_eqs_for_V-full}, we must also take into account the divergency-based equations~\eqref{divD} and \eqref{divB}. Being written in terms of Hertz functions they look as follows,
\begin{subequations}\label{divDdivB->}
\begin{align}
\label{divD_new}
 & \frac{\partial}{\partial\vartheta}
  \left[\varepsilon(\mathbf{r})\frac{\partial V}{\partial\varphi}\right]-
  \frac{\partial}{\partial\varphi}
  \left[\varepsilon(\mathbf{r})\frac{\partial V}{\partial\vartheta}\right]=0\ ,\\[6pt]
\label{divB_new}
 & \frac{\partial}{\partial\vartheta}
  \left[\mu(\mathbf{r})\frac{\partial U}{\partial\varphi}\right]
  -\frac{\partial}{\partial\varphi}
  \left[\mu(\mathbf{r})\frac{\partial U}{\partial\vartheta}\right]=0\ .
\end{align}
\end{subequations}
For the particular case of axially symmetric inclusion in the spherical resonator, where $\partial\varepsilon/\partial\varphi=\partial\mu/\partial\varphi\equiv 0$, equations \eqref{U-third_eq} and \eqref{V-third_eq} get simplified to
\begin{subequations}\label{3d_UV-simpl}
\begin{align}
\label{3d_U-simpl}
  \frac{\partial\varepsilon}{\partial\vartheta}\cdot
  \frac{\partial^2 U}{\partial\varphi\partial r}=0\ ,\\
\label{3d_V-simpl}
  \frac{\partial\mu}{\partial\vartheta} \cdot
  \frac{\partial^2 V}{\partial\varphi\partial r}=0\ ,
\end{align}
\end{subequations}
while equations \eqref{divDdivB->} reduce to equalities
\begin{subequations}\label{dU/d_phi=0|dV/d_phi=0}
\begin{align}
\label{dV/d_phi=0}
 & \frac{\partial\varepsilon}{\partial\vartheta}\cdot
  \frac{\partial V}{\partial\varphi}=0\ ,\\
\label{dU/d_phi=0}
 & \frac{\partial\mu}{\partial\vartheta}\cdot
  \frac{\partial U}{\partial\varphi}=0\ .
\end{align}
\end{subequations}

In the case of central-symmetric non-uniformity, when $\partial\varepsilon/\partial\vartheta=\partial\mu/\partial\vartheta\equiv 0$, equations \eqref{dU/d_phi=0|dV/d_phi=0} become redundant as they do not impose any restrictions on the EM field components. If, however, there appears asymmetry along polar axis, say, if the central points of inner and outer spheres in Figure~\ref{DoubleSpherRes} are shifted by some distance $d\neq 0$, then specific restrictions appear in the form of equalities
\begin{subequations}\label{Restrictions-VU}
\begin{align}
\label{Restrictions-V}
 & H_{\varphi}^m=E_{\vartheta}^m=0\ ,\\
\label{Restrictions-U}
 & E_{\varphi}^e=H_{\vartheta}^e=0\ ,
\end{align}
\end{subequations}
which are valid, respectively, in the regions where $\partial\varepsilon/\partial\vartheta\neq 0$ and $\partial\mu/\partial\vartheta\neq 0$.

Before proceeding with the solution of Eqs.~\eqref{Two_eqs_for_U-full} and \eqref{Two_eqs_for_V-full} it is worth noting that functions $U(\mathbf{r})$ and $V(\mathbf{r})$ directly are not so convenient to represent with them EM fields in a spherical resonator. More convenient for this purpose are Debye potentials \cite{bib:Vainstein88,bib:Oraevskii02}, as indicated above in Eq.~\eqref{U,V-Helm}. The equations for these potentials are more advantageous since, for one thing, they allow to seek the solution of governing equations in the form of expansion in the Fourier series with respect to well-practical spherical functions. One more advantage of using Debye potentials is the possibility to develop for them a constructive perturbation theory allowing to account for different geometrical and electromagnetic properties of inhomogeneities of the resonator infill. That is why we shall continue the exposition in terms of precisely the DPs rather than Hertzian functions.

From here on we will restrict our consideration to the case of non-magnetic resonator infill by putting in all the formulas $\mu(\mathbf{r})\equiv 1$.
Using the uniformity of the resonator dielectric properties in azimuth angle $\varphi$, after transition to Fourier representation with respect to this variable, with Eq.~\eqref{dV/d_phi=0} taken into account, we obtain the following set of control equations for both electric and magnetic Debye potentials,
\begin{subequations}\label{u_eqs-azimuth_m}
\begin{align}
\label{u_eq-azimuth_m}
  m & \left[\hat{\Delta}_m+
  k^2\overline{\varepsilon}-\hat{V}^{(\varepsilon)}-\hat{V}^{(r)}\right]u_m(r,\vartheta)=0\ , \\[6pt]
\label{u_eq-polar_deriv-m}
  \frac{\partial}{\partial\vartheta} &
  \left[\hat{\Delta}_m + k^2\overline{\varepsilon}-\hat{V}^{(\varepsilon)}-
  \hat{V}^{(r)}\right]u_m(r,\vartheta)+
\notag\\
 & \phantom{
  \left[\hat{\Delta}_m + k^2\overline{\varepsilon}-\hat{V}
  \hat{V}^{(r)}\right] }
  +\left\{\frac{\partial}{\partial\vartheta}
  \left[\hat{V}^{(\varepsilon)}+\hat{V}^{(r)}\right]-
  \frac{1}{r^2} \hat{W}_{m}^{(\vartheta)}\right\}u_m(r,\vartheta)=0\ ,
\end{align}
\end{subequations}
\begin{subequations}\label{v_eqs-azimuth_m}
\begin{align}
\label{v_eq_azimuth_m}
  m & \left[\hat{\Delta}_m+
  k^2\overline{\varepsilon}-\hat{V}^{(\varepsilon)}\right]v_m(r,\vartheta)=0\ ,\\
\label{v_eq-polar_deriv-m}
 \frac{\partial}{\partial\vartheta} &
  \left[\hat{\Delta}_m + k^2\overline{\varepsilon}- \hat{V}^{(\varepsilon)}\right]v_m(r,\vartheta)=0\ .
\end{align}
\end{subequations}
Here, $u_m(r,\vartheta)$ and $v_m(r,\vartheta)$ are the $m$-th azimuth Fourier components of the DPs,
\begin{equation}\label{Laplacian-m}
  \hat{\Delta}_m= \frac{1}{r^2}\left[\frac{\partial}{\partial r}\left(r^2\frac{\partial}
  {\partial r}\right)+\frac{1}{\sin\vartheta}\frac{\partial}{\partial\vartheta}
  \left(\sin\vartheta\frac{\partial}{\partial\vartheta}\right)-
  \frac{m^2}{\sin^2\vartheta}\right]
\end{equation}
is the corresponding $m$-th azimuth component of the full Laplace operator, $\overline{\varepsilon}$ is the average (over $r$ and $\vartheta$) value of the permittivity $\varepsilon(\mathbf{r})\equiv \varepsilon(r,\vartheta)$ within the resonator,
\begin{equation}\label{V^(e)-def}
  \hat{V}^{(\varepsilon)}=-k^2\Delta\varepsilon(r,\vartheta)\ ,
\end{equation}
where $\Delta\varepsilon(r,\vartheta)=\varepsilon(r,\vartheta)-\overline{\varepsilon}$, is the local amplitude-type potential accounting for the permittivity heterogeneity,
\begin{equation}\label{V^(r)-def}
  \hat{V}^{(r)}=\frac{\partial\ln\varepsilon(r,\vartheta)}{\partial r}
  \left(\frac{1}{r}+ \frac{\partial}{\partial r} \right)
\end{equation}
is the effective gradient-type potential having the operator nature, whose appearance in equations~\eqref{u_eqs-azimuth_m} is due to the \emph{radial} inhomogeneity of the permittivity, and
\begin{equation}\label{W^(theta)-def}
  \hat{W}_m^{(\vartheta)}=\frac{\partial\ln\varepsilon(r,\vartheta)}{\partial\vartheta}
  \left[\frac{1}{\sin\vartheta}\frac{\partial}{\partial\vartheta}
  \left(\sin\vartheta\frac{\partial}{\partial\vartheta}\right)-
  \frac{m^2}{\sin^2\vartheta}\right]
\end{equation}
is the operator accounting for the \emph{polar-direction} (tangential) non-uniformity of the permittivity function $\varepsilon(r,\vartheta)$.

Equations \eqref{u_eqs-azimuth_m}, \eqref{v_eqs-azimuth_m} which describe TM and TE oscillations, respectively, according to their origination must be satisfied in pairs. Equations \eqref{u_eq-azimuth_m} and \eqref{v_eq_azimuth_m} may be fulfilled either by $m=0$ or by equating to zero the expressions in square brackets. We will consider these two possibilities independently.

%=======================================
\subsection{Zeroth azimuth modes}
\label{0th_azimuth_modes}
%=======================================

For zeroth azimuth modes the first  equations of each of the pairs \eqref{u_eqs-azimuth_m} and \eqref{v_eqs-azimuth_m} are fulfilled automatically. After putting $m=0$ into a pair of other equations, \eqref{u_eq-polar_deriv-m} and \eqref{v_eq-polar_deriv-m}, and using the easily verified commutation relationship
\begin{equation}\label{Delta_0->Delta_1}
  \frac{\partial}{\partial\vartheta}\hat{\Delta}_0=
  \hat{\Delta}_{\pm 1}\frac{\partial}{\partial\vartheta}\ ,
\end{equation}
we obtain the following equation pair,
\begin{subequations}\label{u_0,v_0-eqs}
\begin{align}
\label{u_0-eq}
 & \left[\hat{\Delta}_{\pm 1}+k^2\overline{\varepsilon}
  -\hat{V}^{(\varepsilon)}-\hat{V}^{(r)}-\hat{V}^{(\vartheta)}\right]
  \frac{\partial u_0(r,\vartheta)}{\partial\vartheta}=0\ ,\\
\label{v_0-eq}
 & \left[\hat{\Delta}_{\pm 1}+k^2\overline{\varepsilon}
  -\hat{V}^{(\varepsilon)}\right]\frac{\partial v_0(r,\vartheta)}{\partial\vartheta}=0\ .
\end{align}
\end{subequations}
Here, one more gradient-type effective potential has appeared, viz.,
\begin{equation}\label{V^(theta)-def}
  \hat{V}^{(\vartheta)}=\frac{\partial\ln\varepsilon(r,\vartheta)}{\partial\vartheta}\cdot
  \frac{1}{r^2}\left(\cot\vartheta+\frac{\partial}{\partial\vartheta}\right)\ ,
\end{equation}
which reflects the \emph{eccentricity} property of the resonator infill.

%=======================================
\subsection{Nonzero azimuth modes}
\label{m=/=0_azimuth_modes}
%=======================================

Starting, first, from equations \eqref{v_eqs-azimuth_m}, one can make sure that for $\forall m\neq 0$ both of them are satisfied if the equation is met
\begin{equation}\label{v_m=/=0-eq}
  \left[\hat{\Delta}_m+
  k^2\overline{\varepsilon}-\hat{V}^{(\varepsilon)}\right]v_m(r,\vartheta)=0\ .
\end{equation}
The same does not hold for equations \eqref{u_eqs-azimuth_m}. While the fulfillment of the first of them is provided due to the action of the operator in square brackets, the second one, Eq.~\eqref{u_eq-polar_deriv-m}, reduces to
\begin{equation}\label{u_m=/=0-eq}
  \left\{\frac{\partial}{\partial\vartheta}
  \left[\hat{V}^{(\varepsilon)}+\hat{V}^{(r)}\right]-
  \frac{1}{r^2} \hat{W}_{m}^{(\vartheta)}\right\}u_m(r,\vartheta)=0\ .
\end{equation}
Taking account of definitions \eqref{V^(e)-def}--\eqref{W^(theta)-def} and the differentiation with respect to $\vartheta$ in Eq.~\eqref{u_m=/=0-eq} one can observe that this equation is with certainty satisfied provided that the permittivity is $\vartheta$-uniform, i.\,e., if the equality holds true $\partial\varepsilon(r,\vartheta)/\partial\vartheta\equiv 0$. The opposite case of eccentrically positioned inner sphere requires some extra examination.

Consider the actual case where the permittivity is step-wise dependent on spatial radius-vector, namely, when it is represented by formula
\begin{equation}\label{Epsilon(r)-univ}
  \varepsilon(\mathbf{r})\equiv\varepsilon(r,\vartheta)= \varepsilon_{\mathrm{in}}\bm{\Theta}(\mathbf{r}\in
  \Omega_{\mathrm{in}})+
  \varepsilon_{\mathrm{out}}\bm{\Theta}(\mathbf{r}\in
  \Omega_{\mathrm{out}}\setminus\Omega_{\mathrm{in}})\ ,
\end{equation}
where the notations $\Omega_{\mathrm{in}}$ and $\Omega_{\mathrm{out}}$ are used for the spatial regions enclosed within the inner and outer spheres, respectively; $\bm{\Theta}(A)$ is the logical unit-step function defined as
\begin{equation}\label{bmTheta-def}
  \bm{\Theta}(A)=
 \begin{cases}
  \ 1\ ,& \text{if $A=true$}\ ,\\
  \ 0\ ,& \text{if $A=false$}\ .
 \end{cases}
\end{equation}
For the model \eqref{Epsilon(r)-univ}, the earlier introduced average permittivity is given by
\begin{equation}\label{AverEpsilon}
  \overline{\varepsilon}=
  \frac{\varepsilon_{\mathrm{in}}V_{\mathrm{in}}+
  \varepsilon_{\mathrm{out}}(V_{\mathrm{out}}-V_{\mathrm{in}})}{V_{\mathrm{out}}}=
  \varepsilon_{\mathrm{out}}+
  (\varepsilon_{\mathrm{in}}-\varepsilon_{\mathrm{out}})
  \left(R_{\mathrm{in}}/R_{\mathrm{out}}\right)^3\ ,
\end{equation}
where $V_{\mathrm{in/out}}$ denote the volumes of the inner and outer spheres in Figure~\ref{DoubleSpherRes}.

Due to the presence  of theta-functions in formula~\eqref{Epsilon(r)-univ}, all the terms in
\begin{figure}[h!!]
  \captionstyle{flushleft}
  \centering
  \scalebox{.8}[.8]{\includegraphics{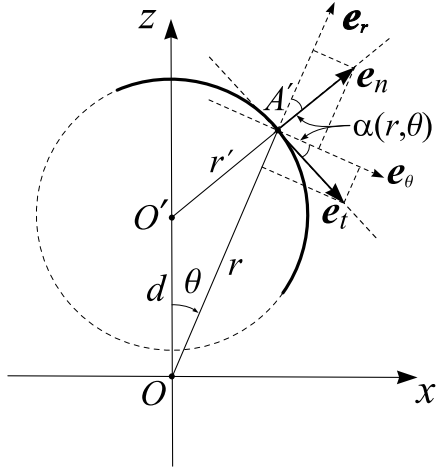}}
  \caption{The diagram for the derivatives calculation.}
\label{fig2}
\end{figure}
Eq.~\eqref{u_m=/=0-eq} have singular nature. To specify their singularities let us express the derivatives with respect to $\vartheta$ and $r$ as the derivatives in the directions normal and tangential to the surface of the inner sphere. In Figure~\ref{fig2}, the corresponding schematics is shown, where $O'$ and $O$ denote the centers of inner and outer sphere, respectively.

For the model \eqref{Epsilon(r)-univ} the derivatives of function $\varepsilon(r,\vartheta)$ with respect to both $r$ and~$\vartheta$ are everywhere equal to zero except for the boundary of the inclusion. At some point $A'$ of the boundary these derivatives are related to the normal and tangential derivatives by formulas
\begin{subequations}\label{DrDO->DnDt}
\begin{align}
\label{Dr->DnDt}
 & \frac{\partial}{\partial r}=
  \cos\alpha(r,\vartheta)\frac{\partial}{\partial l_{n}}-
  \sin\alpha(r,\vartheta)\frac{\partial}{\partial l_{t}}\ ,\\
\label{DO->DnDt}
 & \frac{1}{r}\frac{\partial}{\partial\vartheta}=
  \sin\alpha(r,\vartheta)\frac{\partial}{\partial l_{n}}+
  \cos\alpha(r,\vartheta)\frac{\partial}{\partial l_{t}}\ ,
\end{align}
\end{subequations}
where $\alpha(r,\vartheta)$ is the running angle between the radius-vector of the point $A'$ in primed reference frame and the normal to its non-primed counterpart, see Figure~\ref{fig2}. The tangential derivative of $\varepsilon(r,\vartheta)$ in that point equals to zero while the normal derivative produces a~$\delta$-function. After simple calculus we obtain
\begin{subequations}\label{DrDO->Dn}
\begin{align}
\label{Dr->Dn}
  \left.\frac{\partial\varepsilon(r,\vartheta)}{\partial r}\right|_{A'} &=
  \cos\alpha(r,\vartheta)(\varepsilon_{\mathrm{out}}-\varepsilon_{\mathrm{in}})
  \delta\big[r'(r,\vartheta)-R_{\mathrm{in}}\big]\ ,\\[3pt]
\label{DO->Dn}
  \left.\frac{1}{r}\frac{\partial\varepsilon(r,\vartheta)}{\partial\vartheta}\right|_{A'} &=
  \sin\alpha(r,\vartheta)(\varepsilon_{\mathrm{out}}-\varepsilon_{\mathrm{in}})
  \delta\big[r'(r,\vartheta)-R_{\mathrm{in}}\big]\ .
\end{align}
\end{subequations}

Consider the term in Eq.~\eqref{u_m=/=0-eq} which contains the ``radial-gradient'' potential, $\hat{V}^{(r)}$. The singularity of that term arisen from the double-derivative of function $\varepsilon(r,\vartheta)$ with respect to both of its arguments is compensated by the presence of factor
\begin{equation}\label{Factor_zero}
  \left(\frac{1}{r}+ \frac{\partial}{\partial r} \right)u_m=
  \frac{1}{r}\frac{\partial(ru_m)}{\partial r}=
  \frac{1}{r}\frac{\partial U_m}{\partial r}\ ,
\end{equation}
which for nonzero azimuth modes disappears according to Maxwell equation \eqref{3d_U-simpl}. Two other terms in Eq.~\eqref{u_m=/=0-eq} result in equation
\begin{equation}\label{Unfulfilled_eq_u_m}
  \frac{\partial\varepsilon}{\partial\vartheta}
  \left\{k^2\overline{\varepsilon}-\hat{V}^{(\varepsilon)}+
  \frac{1}{r^2}\left[\frac{1}{\sin\vartheta}\frac{\partial}{\partial\vartheta}
  \left(\sin\vartheta\frac{\partial}{\partial\vartheta}\right)-
  \frac{m^2}{\sin^2\vartheta}\right]\right\}u_m=0\ ,
\end{equation}
which can never be satisfied simultaneously with Eq.~\eqref{u_eq-azimuth_m} except for the cases of either $\partial\varepsilon/\partial\vartheta\equiv 0$ or $u_m\equiv 0$ for $\forall m\neq 0$. This makes us conclude that in the case of eccentrically nonuniform double-spherical resonator only zeroth azimuth mode of TM oscillations is allowed for the existence, while the magnetic-type (TE) oscillations can have any azimuth-mode index.

%=======================================
\section{Separation of TM and TE oscillation modes}
\label{TM-TE_sep}
%=======================================

Although we have uniquely defined the TM and TE oscillations in the resonator under examination, as yet we are not sure if these oscillations can be decoupled from one another, in spite of the fact that they are not directly intermixed in equation~\eqref{u_0-eq} for TM and equations \eqref{v_0-eq} and \eqref{v_m=/=0-eq} for TE oscillation modes. To clarify this position, we must account for boundary conditions (BCs) the fields in our resonator should satisfy.

The main BC for our problem is the vanishing of tangential component of the electric field on the boundary of the resonator, i.\,e., on the outer sphere in Figure~\ref{DoubleSpherRes}. For the geometry chosen this implies the synchronous vanishing of the components $E_{\vartheta}$ and $E_{\varphi}$ of the net electric field at $r=R_{\text{out}}$ at any angles $\vartheta$ and $\varphi$. From Eqs.~\eqref{Ephi_Etheta_Er->U} and \eqref{Hphi_Htheta_Hr->V} it can be seen that the BCs represented in terms of Debye potentials result in the following two equalities,
\begin{subequations}\label{BC(UV)}
   \begin{align}\label{BC(E_theta)}
   \left[\frac{1}{\varepsilon_{\text{out}}\,r}\frac{\partial}{\partial\vartheta}
    \left(u+r\frac{\partial u}{\partial r}\right)+
    \frac{ik}{\sin\vartheta}\frac{\partial v}{\partial\varphi}\right]
    \Bigg|_{r=R_{\text{out}}}=&\ 0\ ,\\
\label{BC(E_phi)}
   \left[\frac{1}{\varepsilon_{\text{out}}\,r\sin\vartheta}\frac{\partial}{\partial\varphi}
    \left(u+r\frac{\partial u}{\partial r}\right)-
    ik\frac{\partial v}{\partial\vartheta}\right]
    \Bigg|_{r=R_{\text{out}}}=&\ 0\ ,
   \end{align}
\end{subequations}
where the fields of TM and TE polarization are seemingly intermixed.

Yet, it is not difficult to observe that equalities \eqref{BC(UV)} would definitely be satisfied if a couple of more simple relations were met, specifically,
\begin{subequations}\label{BC(u|v)}
 \begin{align}\label{BC(Psi_U)}
 \left(u+r\frac{\partial u}{\partial r}\right)\bigg|_{r=R_{\text{out}}}=& 0 \ ,\\[-6pt]
\intertext{and\vspace{-3pt}}
\label{BC(Psi_V)}
 v(r=R_{\text{out}})=& 0 \ .
 \end{align}
\end{subequations}
The choice of the boundary conditions for the above-pointed governing equations, \eqref{u_0,v_0-eqs} and \eqref{v_m=/=0-eq}, in the form of equalities \eqref{BC(u|v)} rather than \eqref{BC(UV)} would allow one to seek the~electric- and magnetic-type fields in the resonator with broken radial and azimuthal symmetries independently of each other, thus representing the oscillations of a~general type as a superposition of \emph{uncoupled} TM and TE oscillation modes.

In the next section we show that the BCs in the form of Eqs.~\eqref{BC(u|v)} are, indeed, admissible for the resonator with inhomogeneous infill, which gives us the opportunity to consider TM and TE oscillations as the uncoupled ones in spite of their being seemingly intermixed by BCs \eqref{BC(UV)}. The proof can be given, however, not in the coordinate representation but rather in the representation of spherical resonator eigenmodes.

%=======================================
\section{Fourier representation of the governing equations}
\label{ModeDecomp}
%=======================================

For the subsequent analysis of working equations \eqref{u_0,v_0-eqs} and \eqref{v_m=/=0-eq} we perform their Fourier transformation from coordinate to momentum representation. For this purpose we will use two different complete eigenfunction sets for the Laplace operator. For a~homogeneous sphere of radius $R_{\mathrm{out}}$ these functions in spherical coordinates may be chosen in the common form
\begin{align}\label{basis_funcs}
 & \ket{\mathbf{r};\bm{\mu}}=\frac{D_n^{(l)}}{R_{\mathrm{out}}}\sqrt{\frac{2}{r}}\,
  J_{l+\frac{1}{2}}\left(\lambda_n^{(l)}r/R_{\mathrm{out}}\right)Y_l^m(\vartheta,\varphi)\\
  \notag
 & \big(n=1,  2,\ldots,\infty\ ;\quad l=0,1,2,\ldots,\infty\ ;\quad  m=-l,-l+1,\ldots l-1,\ l\big)\ ,
\end{align}
where $\bm{\mu}=\{n,l,m\}$ is the vector-valued mode index whose components correspond to radial, polar, and azimuth coordinates, respectively; $D_n^{(l)}$ is the normalization coefficient which depends on the boundary conditions and is represented by the following formulae,
\begin{subequations}\label{ket(r,n)-fin}
\begin{alignat}{3}\label{r_set-Robin}
 &\left[D_n^{(l)}\right]^2  =
   \left[J_{l+\frac{1}{2}}\big(\lambda_n^{(l)}\big)\right]^{-2}
   \left[1-\frac{l(l+1)}{\lambda_n^{(l)}}\right]^{-1}
   & \text{\quad\qquad for BC \eqref{BC(Psi_U)},} \\
 \label{r_set-Dirichlet}
 &\left[D_n^{(l)}\right]^2  =
   \left[J_{l+\frac{3}{2}}\big(\lambda_n^{(l)}\big)\right]^{-2}
   & \text{\quad\qquad for BC \eqref{BC(Psi_V)}.}
\end{alignat}
\end{subequations}
$J_{p}\big(u)$ is the Bessel function of the first kind, $Y_l^m(\vartheta,\varphi)$ is the spherical function; $\lambda_n^{(l)}$ is the set of positive zeros either of the sum $J_{l+\frac{1}{2}}(u)+2uJ'_{l+\frac{1}{2}}(u)$, if boundary condition \eqref{BC(Psi_U)} is applied, or of function $J_{l+\frac{1}{2}}(u)$ in the case of BC \eqref{BC(Psi_V)} (see, e.\,g., \cite{bib:Erdelyi53}, n.~7.10.4). The zeros will be assumed enumerated by natural number $n$ in ascending order. The eigenvalues of the Laplace operator, which correspond to functions \eqref{basis_funcs}, are given by
\begin{equation}\label{Eigen-energies}
  E_{\bm{\mu}}=-k_{\bm{\mu}}^2 =-\left(\frac{\lambda_n^{(l)}}{R_{\mathrm{out}}}\right)^2\ ,
\end{equation}
where the degeneracy in azimuth index $m$ is apparent, whose order is $2l+1$.

As far as we have already done the Fourier transform of working equations over azimuth variable $\varphi$, to complete the transformation of Eqs.~\eqref{u_0,v_0-eqs} and \eqref{v_m=/=0-eq} we will apply the normalized eigenfunction set other than \eqref{basis_funcs}, namely,
\begin{equation}\label{Eigen-funcs_mu_m}
  \ket{\mathbf{r};\bm{\mu}_m}=\frac{D_{n_{\bm{\mu}_m}}^{(l_{\bm{\mu}_m})}}
  {R_{\mathrm{out}}}\sqrt{\frac{2}{r}}\,
  J_{l_{\bm{\mu}_m}+\frac{1}{2}}\left(\lambda_{n_{\bm{\mu}_m}}^{(l_{\bm{\mu}_m})}r/R_{\mathrm{out}} \right) \Theta_{l_{\bm{\mu}_m}m}(\vartheta)\ .
\end{equation}
Here, $\bm{\mu}_m$ is the \emph{two-dimensional} vector mode index, where in the initially three-dimensional index $\bm{\mu}$ the azimuth component is put equal to the specific value $m$,
\begin{equation}\label{Polar_eigen_func}
  \Theta_{lm}(\vartheta)=(-1)^m\left[\frac{2l+1}{2}\cdot\frac{(l-m)!}{(l+m)!}\right]^{1/2}
  P_l^m(\cos\vartheta)
\end{equation}
is the normalized adjoint Legendre function. The set of functions \eqref{Eigen-funcs_mu_m} is orthonormal and complete, with orthonormality and completeness relations having the standard form \cite{bib:Jackson65},
\begin{subequations}\label{Unit_oper_reprs}
\begin{align}
\label{Unit_oper_repr-r}
 & \sum_{\bm{\mu}_m}
  \ket{\mathbf{r};\bm{\mu}_m}\bra{\mathbf{r}';\bm{\mu}_m}=
  \delta(\mathbf{r}-\mathbf{r}')
  \qquad\quad \big(\mathbf{r},\mathbf{r}'\in\Sigma_{\mathrm{out}}\big)\ ,\\
\label{Unit_oper_repr-mu}
 & \int_{\Sigma_{\mathrm{out}}}\!\!\!d\mathbf{r}\,
  \ket{\mathbf{r};\bm{\mu}_m}\bra{\mathbf{r};\bm{\mu}'_m}=
  \delta_{\bm{\mu}_m\bm{\mu}'_m}\ .
\end{align}
\end{subequations}
The summation in Eq.~\eqref{Unit_oper_repr-r} runs over polar and radial mode indices only, $\Sigma_{\mathrm{out}}$ is the outer sphere central cross-section.

The set of eigenfunctions \eqref{Eigen-funcs_mu_m} is convenient in that it is quite universal in form with respect to BCs its constituents may fulfill. We may choose for the Fourier series expansion of functions $u(\mathbf{r})$ and $v(\mathbf{r})$ different basis function sets, being guided by convenience considerations. In particular, in the series
\begin{equation}\label{(u/v)-Fourier}
\begin{pmatrix}
  u(\mathbf{r})\\v(\mathbf{r})
\end{pmatrix}=
  \sum_{\bm{\mu}}
\begin{pmatrix}
  u_{\bm{\mu}}\\v_{\bm{\mu}}
\end{pmatrix}
  \ket{\mathbf{r};\bm{\mu}}\big|_{u,v}
\end{equation}
as a function $\ket{\mathbf{r};\bm{\mu}}\big|_u$ we can take function \eqref{basis_funcs} that meets BC \eqref{BC(Psi_U)}, whereas basis function $\ket{\mathbf{r};\bm{\mu}}\big|_v$ in the expansion of $v(\mathbf{r})$ can be chosen to obey BC \eqref{BC(Psi_V)}. In such a way we, for one thing, fulfill general boundary conditions \eqref{BC(UV)} and, for the other, this permits us to solve the problem for TM and TE oscillations independently of each other. This recipe may be considered as a kind of TM and TE mode separation procedure. The only difference between formulas for electric-type and magnetic-type fields reduces to different sets of zeros $\lambda_n^{(l)}$ entering into the basis functions.

Equations \eqref{u_0,v_0-eqs} and \eqref{v_m=/=0-eq}, which actually define the entire spectrum of the resonator under investigation, are identical in the functional structure and differ only in the set of effective potentials entering into them. To describe the technique of subsequent operations with these equations it is more convenient to work with them not in the initial form but rather to employ the equations for the corresponding Green functions. Such an approach is even more justified if we consider that the oscillation spectrum of the resonator, which in fact is the ultimate goal of our study, is determined by density of states (DOS)
\begin{align}\label{Dens_states->G}
  \nu(k)=\pi^{-1}\mathrm{Im}\big\{\Tr\hat{\mathcal{G}}^{(+)}\big\}\ ,
\end{align}
where $\hat{\mathcal{G}}^{(+)}$ is the Green operator with upper $(+)$ index signifying its retarded nature. By denoting the effective potential, which generally possesses the operator structure, through $\hat{V}$ and accomplishing the Fourier transform of Eqs.~\eqref{u_0,v_0-eqs} and \eqref{v_m=/=0-eq} with respect to coordinate variables $r$ and $\vartheta$ we arrive at the following infinite set of coupled algebraic equations for the Green operator mode matrix elements, viz.,
\begin{equation}\label{GF_eqs-schemat}
  \big(k^2\overline{\varepsilon}-k_{\bm{\mu}_m}^2
  -\mathcal{V}_{\bm{\mu}_m}\big)G_{\bm{\mu}_m\bm{\mu}'_m}
  -\sum_{\bm{\nu}_m\neq\bm{\mu}_m}
  \mathcal{U}_{\bm{\mu}_m\bm{\nu}_m}G_{\bm{\nu}_m\bm{\mu}'_m}=
  \delta_{\bm{\mu}_m\bm{\mu}'_m}\ .
\end{equation}
Matrix element
\begin{equation}\label{Oper_pots_matr_els}
  \mathcal{U}_{\bm{\mu}_m\bm{\nu}_m}=
  \int_{\Sigma_{\mathrm{out}}}\!\!\!d\mathbf{r}\,
  \bra{\mathbf{r};\bm{\mu}_m}\hat{V}\ket{\mathbf{r};\bm{\nu}_m}
\end{equation}
of the effective potential, in view of the restriction $\bm{\nu}_m\neq\bm{\mu}_m$ in the sum of Eq.~\eqref{GF_eqs-schemat}, will be henceforth referred to as the \emph{inter-mode} effective potential, while the diagonal element of the potential matrix, $\mathcal{V}_{\bm{\mu}_m}\equiv\mathcal{U}_{\bm{\mu}_m\bm{\mu}_m}$, which simply renormalizes energetic parameter $k_{\bm{\mu}_m}^2$ without change in the mode index, will be termed the \emph{intra-mode} potential.

In the further analysis we will regard the average dielectric constant of the resonator as either purely real or complex-valued quantity, considering either a dissipation-free theoretical model or realistic systems with dissipation. In the latter case we will assume that $\overline{\varepsilon}=\overline{\varepsilon}'+i\overline{\varepsilon}''$, where $\overline{\varepsilon}'$ and $\overline{\varepsilon}''$ denote the real and imaginary parts of the average permittivity, respectively. In general, the specific value of $\overline{\varepsilon}''$ is dependent on both of the dissipative parts of $\varepsilon_{\mathrm{in}}$ and $\varepsilon_{\mathrm{out}}$, which even may have different signs. Yet, everywhere below we will assume that $\overline{\varepsilon}''\geqslant 0$.

The set of equations \eqref{GF_eqs-schemat}, which is identical in form for both TM and TE oscillations, can be solved either numerically or analytically. The numerical calculations, although they, in principle, can be carried out for any value of dissipation parameter $\overline{\varepsilon}''$, are quite time consuming for quasioptical resonators, especially in the case where dissipation is  taken finite. The latter case, however, can be effectively analyzed by analytical means, if one uses the method of effective disconnection of the resonance modes of an~inhomogeneous system, which is detailed in Ref.~\cite{bib:GanErTar07}. Here we will give a brief sketch of the method and present the result of its implementation.

%=======================================
\section{Analytic solution of equation set (\ref{GF_eqs-schemat})}
\label{Operat_sol}
%=======================================

Considering equations \eqref{GF_eqs-schemat} one can observe that if all the inter-mode potentials were equal to zero the Green function matrix would be essentially diagonal,
\begin{equation}\label{Gmumu'-unpert}
   G_{\bm{\mu}_m\bm{\mu}'_m}=G_{\bm{\mu}_m}^{(V)}\delta_{\bm{\mu}_m\bm{\mu}'_m}\ .
\end{equation}
Here, function $G_{\bm{\mu}_m}^{(V)}$, which we term henceforth the \emph{trial} mode Green function, is given by
\begin{equation}\label{Trial_G_mu_m}
  G_{\bm{\mu}_m}^{(V)}(k)  = \left( k^2\overline{\varepsilon}-k_{\bm{\mu}_m}^2 - {\cal V}_{\bm{\mu}_m} \right)^{ - 1}\ .
\end{equation}
This function already contains some information regarding the cavity non-uniformity, yet this information is only partial, concentrated in the intra-mode potential ${\cal V}_{\bm{\mu}_m}$. The inter-mode potentials in Eq.~\eqref{GF_eqs-schemat} result in effective ``dynamic'' inelasticity of the trial mode scattering, which makes actual states of the field in the cavity ill defined. This kind of inelasticity is due to the failure of the wave phase in the course of its propagation in the non-uniform medium, and it has little in common with true inelasticity due to the absorption or gain.

By simple algebraic manipulations (see Ref.~\cite{bib:GanErTar07} for details), from the set of equations \eqref{GF_eqs-schemat} the linear connection between non-diagonal and diagonal elements of the mode Green matrix results, which in the operator form is given~by
\begin{equation}\label{Gnu->Gmumu_m}
  G_{\bm{\nu}_m\bm{\mu}_m}=\hat{\mathbf{P}}_{\bm{\nu}_m}
  (\hat{\mathbf{1}}-\hat{\mathsf R}_m)^{-1}\hat{\mathsf R}_m
  \hat{\mathbf{P}}_{\bm{\mu}_m}G_{\bm{\mu}_m\bm{\mu}_m}\ .
\end{equation}
Here, $\hat{\mathbf{P}}_{\bm{\mu}_m}$ is the Feshbach-type projection operator \cite{bib:Feshbach58,bib:Feshbach62} whose action reduces to the assignment of the value $\bm{\mu}_m$ to the nearest mode index of any adjacent operator, no matter where it may stand --- to the left or to the right of $\hat{\mathbf{P}}_{\bm{\mu}_m}$; $\hat{\mathsf{R}}_m=\hat{\mathcal{G}}_m^{(V)}\hat{\mathcal{U}}_m$ is the \emph{mode-mixing operator} defined on the reduced space of mode indices, $\mathsf{\overline M}_{\bm{\mu}_m}$, which includes all the indices of the resonator modes save index $\bm{\mu}_m$; $\hat{\mathcal{G}}_m^{(V)}$ is the operator with the diagonal matrix consisting of trial Green functions~\eqref{Trial_G_mu_m}, $\hat{\mathcal{U}}_m$ is the operator with null-diagonal matrix, which incorporates all the inter-mode potentials entering into Eq.~\eqref{GF_eqs-schemat}. Assuming then in equation~\eqref{GF_eqs-schemat} $\bm{\mu}'_m=\bm{\mu}_m$ and substituting into the achieved equation all the intermode propagators in the form \eqref{Gnu->Gmumu_m} we arrive at the following expression for the exact intramode propagator~$G_{\bm{\mu}_m\bm{\mu}_m}$, viz.,
\begin{equation}\label{Gmumu-last}
  G_{\bm{\mu}_m\bm{\mu}_m}=\left[k^2\overline{\varepsilon}-
  k^2_{\bm{\mu}_m}-\mathcal{V}_{\bm{\mu}_m}-\mathcal{T}_{\bm{\mu}_m}(k)\right]^{-1}\ .
\end{equation}
Here,
\begin{equation}\label{T-oper}
  \mathcal{T}_{\bm{\mu}_m}(k)=\hat{\mathbf{P}}_{\bm{\mu}_m}\hat{\mathcal{U}}_m
  (\hat{\mathbf{1}}-\hat{\mathsf{R}}_m)^{-1}\hat{\mathsf{R}}_m\hat{\mathbf{P}}_{\bm{\mu}_m}
\end{equation}
is the portion of the mode $\bm{\mu}_m$ eigen-energy which is related to the intermode scattering only. The validity condition for the expressions \eqref{T-oper} and \eqref{Gnu->Gmumu_m} is the non-singularity of the inverse operator entering into them. To meet this requirement, it suffices for the operator $\hat{\mathsf{R}}_m$ not to have the unity among its eigenvalues. The universal method to accomplish this task is to make the wave operator in Eqs.~\eqref{u_0,v_0-eqs} and \eqref{v_m=/=0-eq} non-Hermitian, just like in the papers by Feshbach \cite{bib:Feshbach58,bib:Feshbach62}. However, if there it was naturally achieved due to the openness of the system, in our case the only suitable way is to add to the average permittivity of the resonator infill some imaginary part related to dissipation or gain, whatever small. In the final results this imaginary addendum can be directed to zero if there is a need to consider the dissipation-free system.

The main feature of expression \eqref{Gmumu-last} for the Green matrix diagonal element is that effective potential $\mathcal{T}_{\bm{\mu}_m}(k)$, which is known as \emph{optical potential} in the quantum theory of scattering \cite{bib:WuOmura62}, depends in a quite complicated manner on~$k$, which is the consequence of the dependence on frequency of the trial Green functions entering into the mode-mixing operator $\hat{\mathsf{R}}_m$. Typically, optical potentials are complex-valued, if it comes to open systems. Yet, we study the conservative system, so the complexity necessary for the inverse operator in Eq.~\eqref{T-oper} to be properly defined is introduced through the seed dissipation or gain hidden in the imaginary part of the infill permittivity. Even infinitesimal value of $\varepsilon''$ can result in difficult-to-control value of $\mathrm{Im}\,\mathcal{T}_{\bm{\mu}}$, of which one can make sure by rewriting potential \eqref{T-oper} as
\begin{align}\label{Tpot->matr_el-fin}
  \mathcal{T}_{\bm{\mu}}(k) &=\sum_{\bm{\nu}_1,\bm{\nu}_2\neq\bm{\mu}}
  \mathcal{U}_{\bm{\mu}\bm{\nu}_1}
  \left\{\left[\hat{\mathbf{1}}-
  \hat{\mathcal{G}}^{(V)}\hat{\mathcal{U}}\right]^{-1}\right\}_{\bm{\nu}_1\bm{\nu}_2}
  \left(\hat{\mathcal{G}}^{(V)}\hat{\mathcal{U}}\right)_{\bm{\nu}_2\bm{\mu}}
  \notag\\*
  &=
  \sum_{\bm{\nu}_1,\bm{\nu}_2\neq\bm{\mu}}
  \mathcal{U}_{\bm{\mu}\bm{\nu}_1}
  \left[\left(\triangle+k^2\overline{\varepsilon}-\hat{\mathcal{V}}-
  \hat{\mathcal{U}}\right)^{-1}\right]_{\bm{\nu}_1\bm{\nu}_2}
  \mathcal{U}_{\bm{\nu}_2\bm{\mu}}
  \notag\\*
  &=\sum_{\bm{\nu}_1,\bm{\nu}_2\neq\bm{\mu}}
  \mathcal{U}_{\bm{\mu}\bm{\nu}_1}
  \left[\left(\triangle+k^2\overline{\varepsilon}^{\,*}-\hat{\mathcal{V}}-
  \hat{\mathcal{U}}\right)\left|\left(\triangle+k^2\overline{\varepsilon}-\hat{\mathcal{V}}-
  \hat{\mathcal{U}}\right)\right|^{-2}\right]_{\bm{\nu}_1\bm{\nu}_2}
  \mathcal{U}_{\bm{\nu}_2\bm{\mu}}
\end{align}
(to be short, we omit the insignificant index $m$). It can be seen that equality $\mathrm{Im}\,\mathcal{T}_{\bm{\mu}}=0$ implies the quantity $\overline{\varepsilon}$ to have exactly zero imaginary part. In the opposite case, the value of $\mathrm{Im}\,\mathcal{T}_{\bm{\mu}}$ is highly dependent on $k$, which results in different local widths of the resonances.

If one would go to find spectral values of $k$ through the poles and other exceptional points of Green function \eqref{Gmumu-last}, in view of extreme complexity of $\mathrm{Re}\,\mathcal{T}_{\bm{\mu}_m}(k)$ the available values of $k$ were not all real. The situation here is reminiscent of that considered by Hatano in studies on open quantum-mechanical systems \cite{bib:Hatano13}. In our case the effective openness appears if one turns to the trajectory-based interpretation of resonant states in the cavity under consideration \cite{bib:Gutzwiller71}. Thus arising the finite width of the spectral lines can lead to their significant overlap, which may be interpreted as the interaction of resonances declared to be the essential source of spectral chaos \cite{bib:ZaslChir72}.

%=======================================
\section{Numerical verification of the developed theory}
\label{Num_anal}
%=======================================

Unfortunately, in the general case the above developed method does not allow obtaining the exact analytic expressions for the eigen-frequencies of an eccentrically layered spherical cavity resonator. To overcome this disadvantage by numerical methods, we have developed a software that implements the proposed theory. The software uses the well-known library for scientific computing~\cite{bib:Galassi03} and is based on the MPI parallel computing technology. The calculations were performed on the grid cluster of the Institute for Radiophysics and Electronics of NAS of Ukraine.

\begin{figure}[h!!]
  \captionstyle{normal}
  \centering
  \scalebox{.3}[.3]{\includegraphics{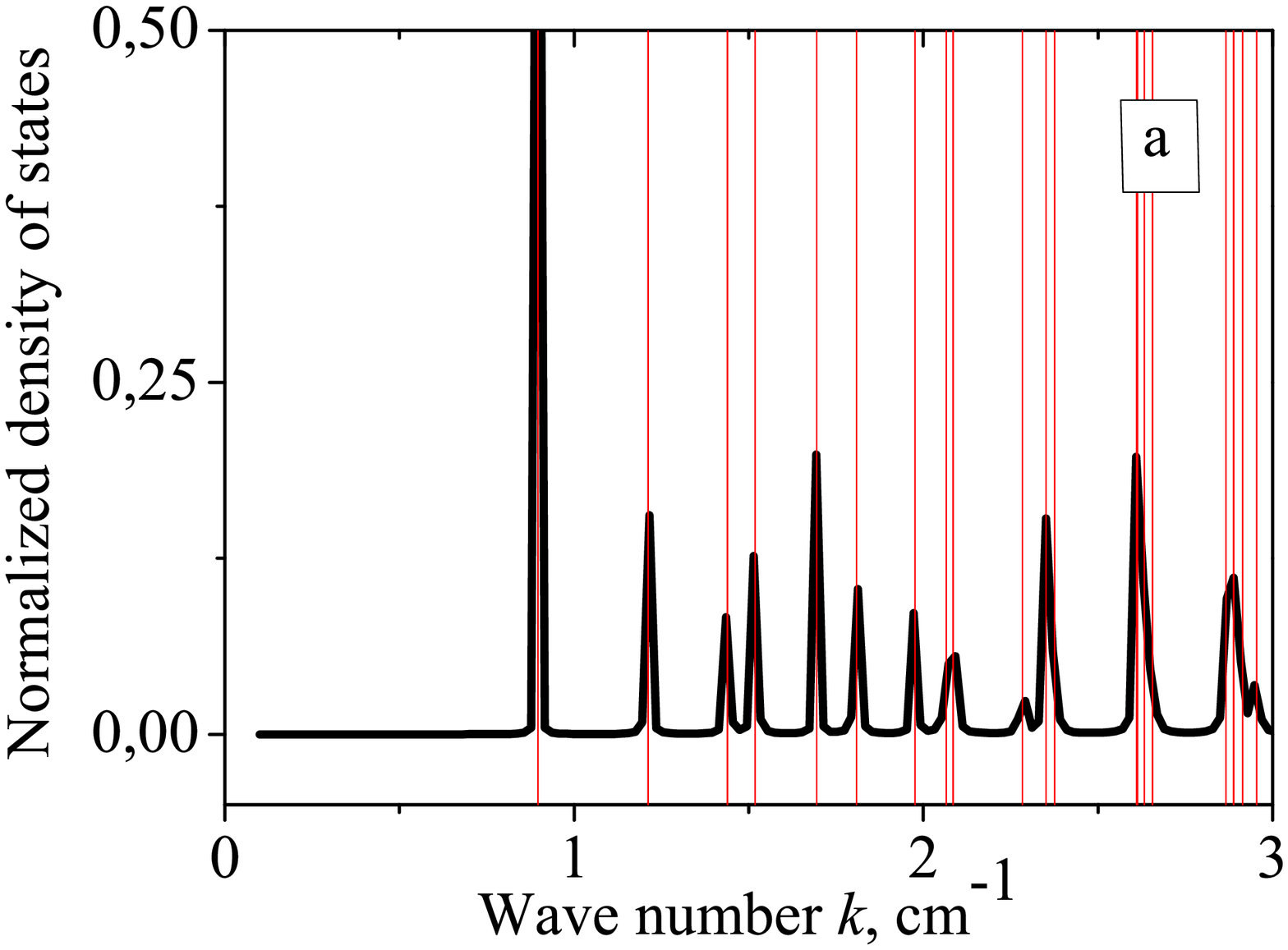}
  \includegraphics{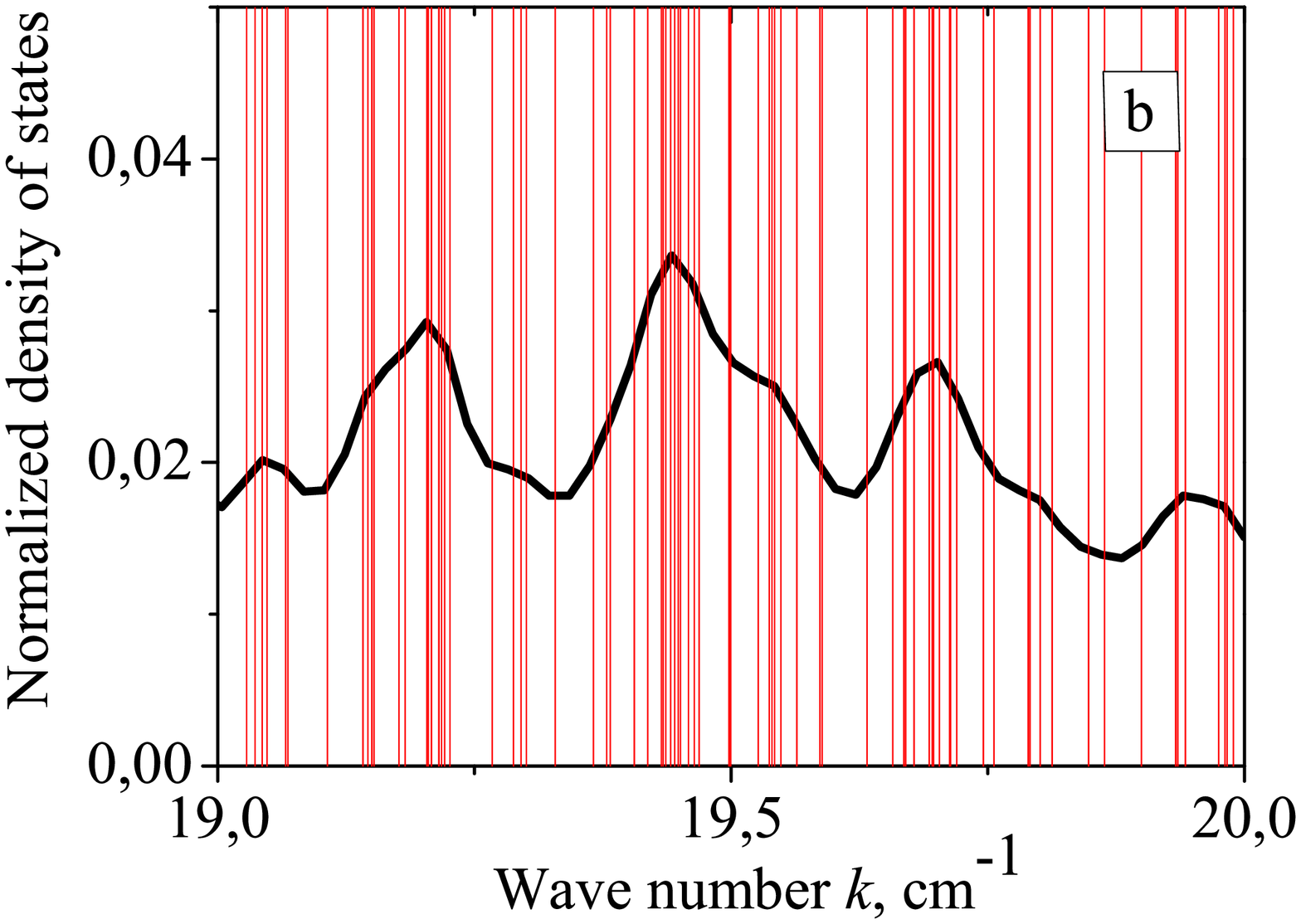}}
  \caption{The normalized density of states (black curves) and the matrix eigenvalues (red vertical lines) for TE oscillations of the concentrically layered spherical cavity in two particular frequency domains. The resonator parameters are taken as follows: $R_{\mathrm{in}}=2$~cm, $R_{\mathrm{out}}=4.46$~cm, $\varepsilon_{\mathrm{in}}= 2.08+0.004j$, $\varepsilon_{\mathrm{out}}= 1+0.004j$, $d = 0$; 1000 basis modes were used when solving matrix equation~\eqref{GF_eqs-schemat}.}
\label{fig3}
\end{figure}

To verify the proposed theory and the software developed, the eigen-frequencies of the central symmetric layered cavity resonator (whose spectrum can be found by other known methods) were calculated. For comparison, we also used the method of solving the Maxwell equations in the outer and inner layers with subsequent matching the respective solutions at the inner boundary of the resonator. The results are presented in Figure~\ref{fig3}, where we provide two different parts of the spectrum of spherical resonator with centralized spherical insert of a given radius.
Thin vertical lines depict the positions of the expected resonance maxima of the Green function, which are found numerically from Eq.~\eqref{GF_eqs-schemat} with $\varepsilon''$ from the outset taken exactly zero.

By the solid lines in Figure~\ref{fig3}, the spectrum obtained from Eq.~\eqref{Dens_states->G} with Green matrix elements taken in the form~\eqref{Gmumu-last} is shown, where the dissipation is supposed to be finite but vanishingly small. It is evident that with the presence of dissipation/gain mechanisms in the system the spectrum becomes highly rarefied due to mutual absorption of the smoothed resonances. Moreover, the spectrum, being (quasi-)discrete for the resonator of simple integrable form, tends to develop into the continuous one when the ray dynamics comes to be to a certain degree non-integrable. A significant overlap of the resonances occurs even at infinitesimal value of dissipation, which stems from the fact that the ray trajectories in the resonator with broken symmetry, no matter how much it would be, after a sufficiently large period of time evolve into chaotic bundles. According to the widespread Gutzwiller's periodic-orbit theory \cite{bib:Gutzwiller71}, the bursts on the continuous spectral line correspond to the most stable and long-lived quasi-classic trajectories while the smooth spectral ``background'' fits with the set of chaotic trajectory sets. We will discuss in detail the chaotic properties of the spectrum in the subsequent publication, which we intend to devote to the statistical analysis of the inter-frequency intervals extracted from the above obtained formulas.

\begin{figure}[h]
 \captionstyle{normal}
 \centering
 \scalebox{.3}[.3]{\includegraphics{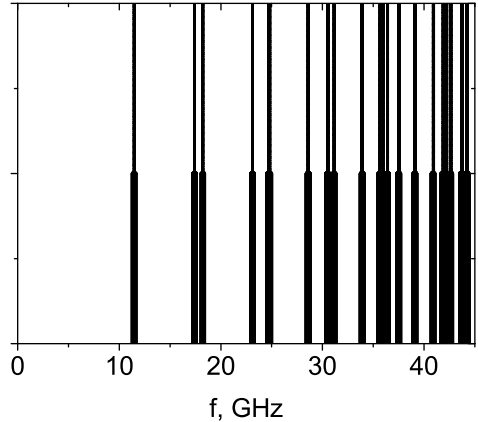}}
 \caption{The comparison of the frequency spectrum of the layered central symmetric dielectric cavity. Thin lines result from the dissipation-free version of equation set \eqref{GF_eqs-schemat} whereas thick lines follow from the CST Microwave Studio simulations with application of the Eigen mode solver. The resonator parameters are as follows: $R_{\mathrm{in}}=5$~mm, $R_{\mathrm{out}}=10$~mm, $\varepsilon_{\mathrm{in}}= 2.08$, $\varepsilon_{\mathrm{out}}= 1$, $d = 0$. }
\label{fig4}
\end{figure}

For additional verification, we find the limited set of eigen-frequencies of the central symmetric layered cavity by means of the dissipation-free version of equation \eqref{GF_eqs-schemat} and in parallel got them from the CST Microwave Studio simulations with the use of the Eigen mode solver. A comparison of the results of these two independent methods is presented in Figure~\ref{fig4}, wherefrom their coincidence is certain.

\begin{figure}[h!!]
  \captionstyle{normal}
\begin{minipage}[b]{0.9\linewidth}
  \centering
\hspace{1cm}  \scalebox{.4}[.4]{\includegraphics{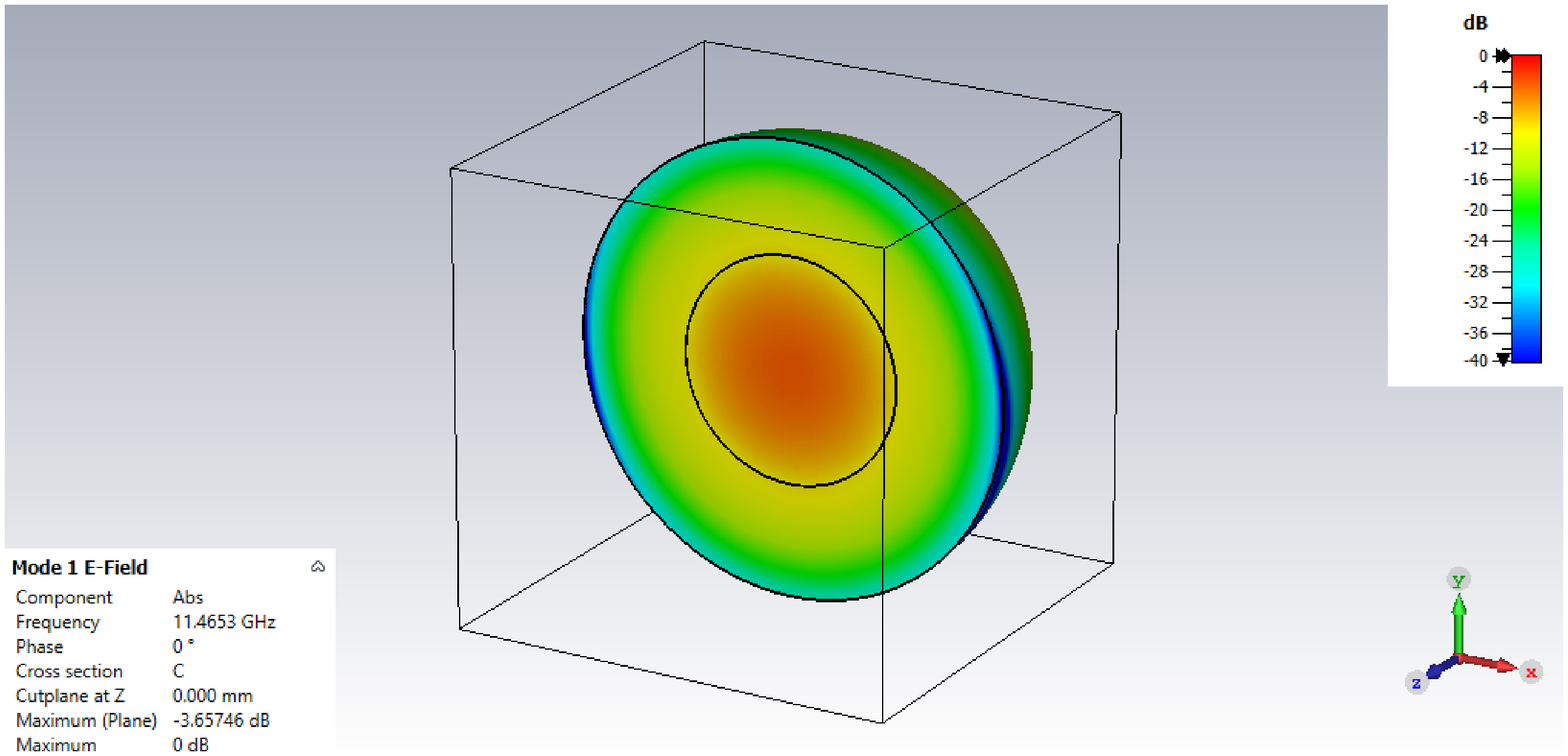}} \\[-.5\baselineskip]
\hspace{1cm} \text{\footnotesize \textit{(a) $|\textbf{E}\,|$ distribution}}
\end{minipage} \\[.5\baselineskip]
\begin{minipage}[b]{0.9\linewidth}
  \centering
\hspace{1cm}  \scalebox{.4}[.4]{\includegraphics{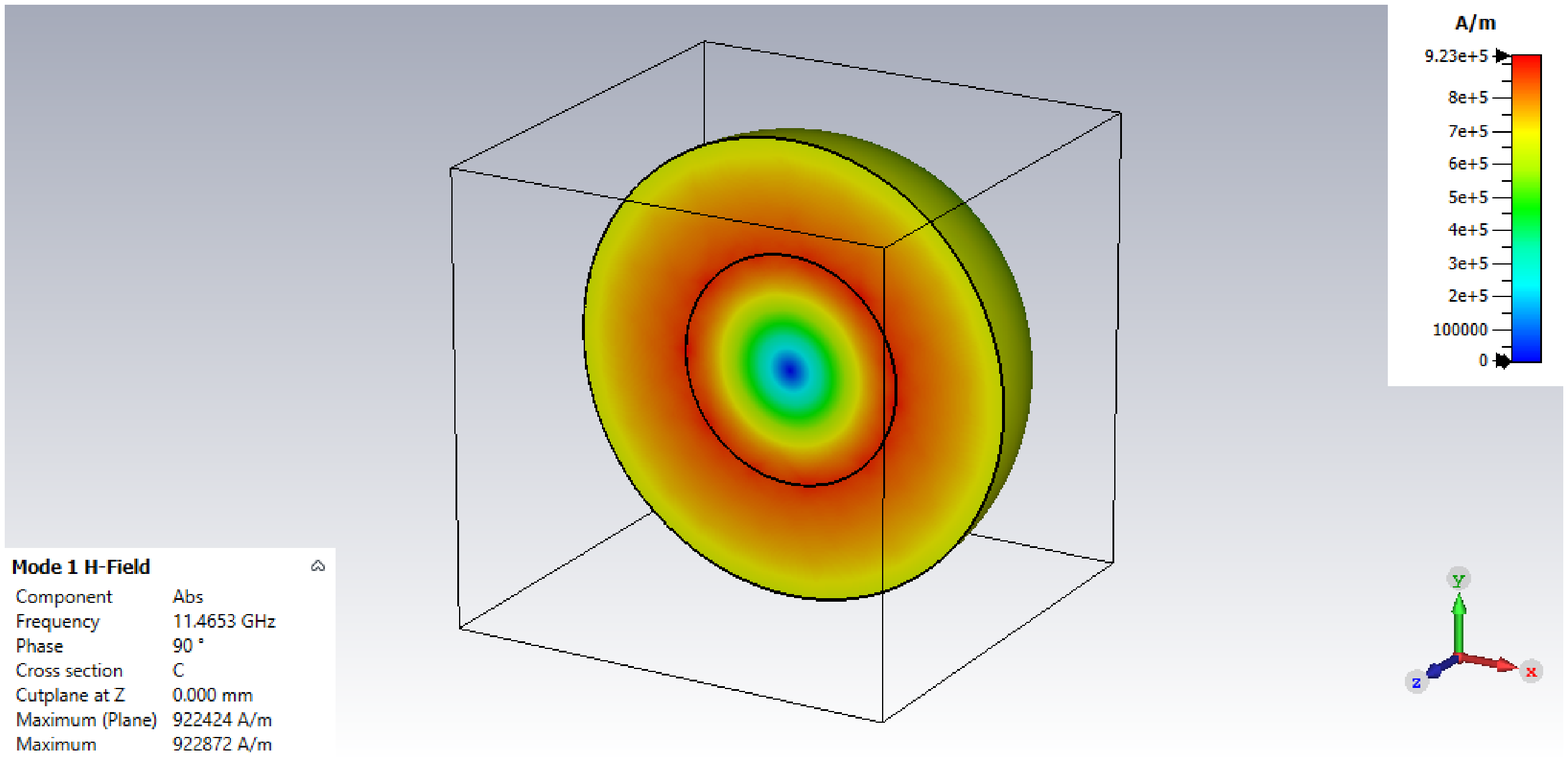}} \\[-.5\baselineskip]
\hspace{1cm}  \text{\footnotesize \textit{(b) $|\textbf{H}\,|$ distribution}}
\end{minipage}
  \caption{Distribution of the absolute values of electric \textit{(a)} and magnetic \textit{(b)} fields for the lowest-frequency mode (11.47 GHz); cut plane through the cavity center. The resonator parameters are the same as in Figure~\ref{fig4}.}
\label{fig5}
\end{figure}

Additionally, in Figure~\ref{fig5} we present the distribution of electromagnetic field of the first mode of the eigen- mode set obtained using CST Microwave Studio simulations. This resonant mode, as expected, is uniform in the azimuth angle $\varphi$ for both the electric and magnetic fields.
%

%=======================================
\section{Concluding remarks}
\label{Concl_remarks}
%=======================================

In the present paper we present a theory that is capable of the solution to Maxwell equations in spherical cavity resonators with non-homogeneous dielectric infill. Specifically, we have examined the resonator whose non-uniformity is due to a spherical dielectric insert with constant permittivity different from that of the host resonator. The position of the insert inside the host uniform cavity as well as its radius are assumed to be optional. To obtain the oscillation spectrum, the vector Maxwell equations are reduced to the set of two scalar equations for Hertzian functions of electric and magnetic types, which have the form of Helmholtz equations subject to the perturbation. The similarity of these equations to Schr\"odinger equation has made it possible to avoid their solving in different parts of the resonator, with necessary joining of the solutions at the boundaries between them. We choose the method to solve the equations within the whole resonator by representing its heterogeneity in terms of effective potentials, which appear to be of two types --- the amplitude-type and the gradient-type potentials.

The EM field in the angularly symmetric cavity can always be represented in terms of TM and TE Fourier components. In this case the general solution is shown to be a superposition of TM and TE oscillations which are disentangled in spite of the particular radial non-uniformity. If the resonator is angularly symmetric, whatever the non-regularity in radial direction, there are present all angular modes in its spectrum, both of TM and TE type. Each type of the oscillations is described by only one scalar effective potential, which substantially reduces the computational resources required for finding the resonator spectrum by means of computer simulations.

The violation of angular symmetry in the infill permittivity dramatically affects the spectrum, namely, the structure and the number of degrees of freedom (independent variables) of the EM field in the cavity. Misalignment of the insert with regard to the host resonator center makes all azimuth modes of TM oscillations, except for the zeroth (uniform) mode, become forbidden in the resonator in question, while TE-polarized oscillations are allowed to have any index of the azimuthal mode. The asymmetry of this particular kind is related to the fact that in the present study the permittivity only is chosen to be non-uniform while the permeability is taken to be everywhere constant. In case of the both electrically and magnetically non-smooth axially symmetric resonator with eccentric infill, only azimuthally uniform eigenmodes can be represented as the independent TE and TM oscillations, i.\,e., by the pair of uncoupled scalar Hertz potentials. Other azimuthal modes, even if they exist in the spectrum, cannot be found in frameworks of the method presented here and thus should be categorized as the hybrid modes.

The spectrum of non-homogeneous resonator, no matter angularly symmetric or not, is shown to acquire some chaotic properties which cannot be meaningfully analyzed without invoking the specific methods of the chaos theory. The detailed analysis of this issue will be presented in the subsequent publication.

%%%%%%%%%%%%%%%%%%%%%%%%%%%%%%%%%%%%%%%
\appendix
%%%%%%%%%%%%%%%%%%%%%%%%%%%%%%%%%%%%%%%

%=======================================
\section{Evaluation and comparison of effective potentials in E\lowercase{q}.~(\ref{GF_eqs-schemat})}
\label{Eff_pots_eval}
%=======================================

{\allowdisplaybreaks
The effective potential $\hat{V}$ in Eq.~\eqref{Oper_pots_matr_els} is, in general, a sum of three terms, $\hat{V}^{(\varepsilon)}$, $\hat{V}^{(r)}$ and $\hat{V}^{(\vartheta)}$, of which the first term enters both into Eqs.~\eqref{u_0,v_0-eqs} and \eqref{v_m=/=0-eq} while the other two into Eq.~\eqref{u_0-eq} only.

Consider, first, the potential $\hat{V}^{(\varepsilon)}$. As opposed to gradient potentials $\hat{V}^{(r)}$ and $\hat{V}^{(\vartheta)}$, the amplitude-type potential $\hat{V}^{(\varepsilon)}$ is not quite perceptive to the coordinate system one chooses for its calculation. For the purposes of numerical calculations it is preferable to take for this potential the coordinate system $K$ located at the center of outer sphere, see the diagram in Figure~\ref{fig6}.
\begin{figure}[h]
 \captionstyle{flushleft}
 \centering
 \scalebox{.5}[.5]{\includegraphics{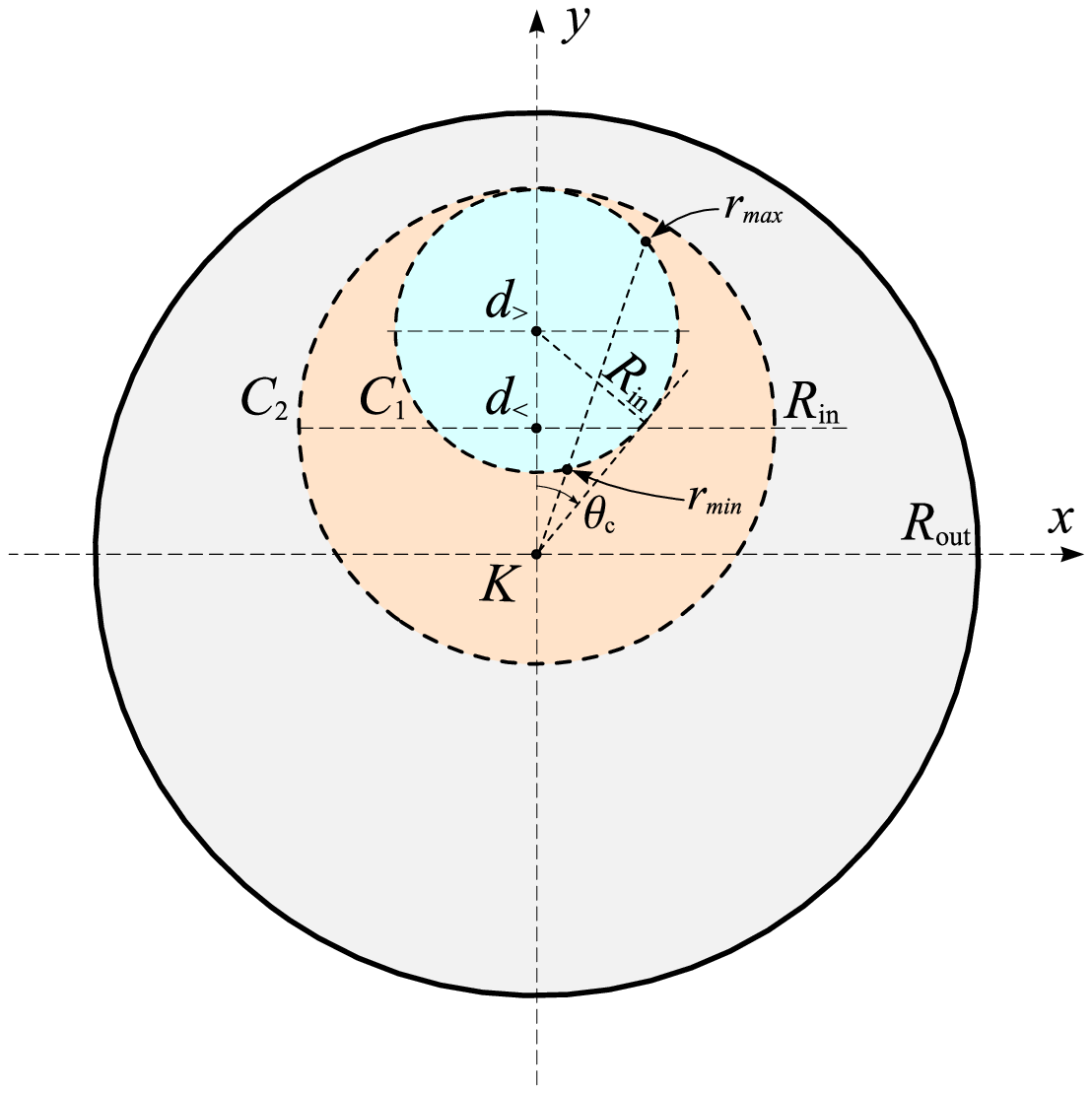}}
 \caption{To the derivation of formula \eqref{Int_IN(e)} in the $K$ frame. Symbols $d_{\gtrless}$ in the figure correspond to $d\gtrless R_{\mathrm{in}}$.}
%\label{D_Sphere-aux1}
\label{fig6}
\end{figure}
Taking account of piecewise continuity of the resonator infill, the mode matrix element of $\hat{V}^{(\varepsilon)}$ can be represented as
\begin{equation}\label{U_munu^(delta_epsilon)-matr_elem}
  \mathcal{U}^{(\varepsilon)}_{\bm{\mu}_m\bm{\nu}_m}=
  -k^2(\varepsilon_{\mathrm{in}}-\varepsilon_{\mathrm{out}})
  \Big[I_{\bm{\mu}_m\bm{\nu}_m}^{(\varepsilon)}-
  \big(R_{\mathrm{in}}/R_{\mathrm{out}}\big)^3\delta_{\bm{\mu}_m\bm{\nu}_m}\Big]\ ,
\end{equation}
where
\begin{align}\label{Int_IN(e)}
  I_{\bm{\mu}_m\bm{\nu}_m}^{(\varepsilon)}=& \int\limits_{\Sigma_{\mathrm{in}}}d\mathbf{r}
  \langle\mathbf{r};\bm{\mu}_m|\mathbf{r};\bm{\nu}_m\rangle =
  \frac{2}{R^2_{\mathrm{out}}}D_{n_{\bm{\mu}_m}}^{(l_{\bm{\mu}_m})}
  D_{n_{\bm{\nu}_m}}^{(l_{\bm{\nu}_m})}
  \times
  \notag\\*
  &
  \times
    \int\limits_{\Sigma_{\mathrm{in}}}
  \frac{d\mathbf{r}}{r}
  J_{l_{\bm{\mu}_m}+\frac{1}{2}}\left(\lambda_{l_{\bm{\mu}_m}}^{(l_{\bm{\mu}_m})}r/R_{out}\right)J_{l_{\bm{\nu}_m}+\frac{1}{2}}\left(\lambda_{l_{\bm{\nu}_m}}^{(l_{\bm{\nu}_m})}r/R_{out}\right)
  \Theta_{l_{\bm{\mu}_m}m}(\vartheta)\Theta_{l_{\bm{\nu}_m}m}(\vartheta)\ ,
\end{align}
$\Sigma_{\mathrm{in}}$ is the inner sphere central cross-section.

Transform the integral in \eqref{Int_IN(e)} for two variants of the inner sphere, encircling and not encircling the center of the outer sphere. Consider, first, the inner sphere centered at point $d_>$ in Figure~\ref{fig6}, when the point $K$ does not fall into it. The integral over region $\Sigma_{\mathrm{in}}$ reduces to integration with respect to $\vartheta$ from zero to $\vartheta_{c}=\arcsin(R_{\mathrm{in}}/d_>)$ while the integral with respect to $r$ runs in the interval $r_{min}(\vartheta)<r<r_{max}(\vartheta)$, with $r_{min}(\vartheta)$ and $r_{max}(\vartheta)$ being the points where the ray going from $K$ intersects the circle~$C_1$. As far as this circle is described by equation
\begin{equation}\label{C_1-spherical}
  r^2-2rd_>\cos\vartheta+d_>^2=R_{\mathrm{in}}^2\ ,
\end{equation}
one can  easily find that
\begin{equation}\label{r_min/max}
  r_{\genfrac{}{}{0pt}{3}{max}{min}}(\vartheta)\big|_{C_1}=d_>\cos\vartheta\pm
  \sqrt{R_{\mathrm{in}}^2-d_>^2\sin^2\vartheta}\ .
\end{equation}

If the coordinate origin $K$ lies within the inner sphere, integration with respect to $\vartheta$ in \eqref{Int_IN(e)} is done from $0$ to $\pi$ while integration with respect to $r$ runs from $0$ to $r_{max}(\vartheta)$. In the general case the expression for $I_{\bm{\mu}_m\bm{\nu}_m}^{(\varepsilon)}$ can be written in the universal form
\begin{align}
\label{Int_INTER-aux_1}
  I_{\bm{\mu}_m\bm{\nu}_m}^{(\varepsilon)}=
  \frac{B_{\bm{\mu}_m\bm{\nu}_m}}{R_{\mathrm{out}}^2}
 & \int\limits_{t_{min}}^{1} dt
  P_{l_{\bm{\mu}_m}}^{m}(t)
  P_{l_{\bm{\nu}_m}}^{m}(t)
  \times
  \notag\\*
  &
  \times
  \int\limits_{r_{min}(t)}^{r_{max}(t)}rdr  J_{l_{\bm{\mu}_m}+\frac{1}{2}}\left(\lambda_{n_{\bm{\mu}_m}}^{(l_{\bm{\mu}_m})}r/R_{\mathrm{out}}\right)
  J_{l_{\bm{\nu}_m}+\frac{1}{2}}\left(\lambda_{n_{\bm{\nu}_m}}^{(l_{\bm{\nu}_m})}r/R_{\mathrm{out}}\right)\ ,
\end{align}
where the notations are used
\begin{subequations}\label{theta_c,r_minmax}
\begin{align}
\label{theta_c}
 & t_{min} =\theta(d-R_{\mathrm{in}})\sqrt{1-(R_{\mathrm{in}}/d)^2}
 -\theta(R_{\mathrm{in}}-d)\ , \\
\label{r_min(theta)}
 & r_{min}(t)=\theta(d-R_{\mathrm{in}})
  \Big[dt - \sqrt{R_{\mathrm{in}}^2-d^2(1-t^2)}\Big]\ ,\\
\label{r_max}
 & r_{max}(t)=dt +
  \sqrt{R_{\mathrm{in}}^2-d^2(1-t^2)}\ ,
\end{align}
\end{subequations}
$\theta(\ldots)$ is the Heaviside unit step function, and the coefficient $B_{\bm{\mu}_m\bm{\nu}_m}$ has the form
\begin{equation}\label{B_nlm}
  B_{\bm{\mu}_m\bm{\nu}_m}=
  D_{n_{\bm{\mu}_m}}^{(l_{\bm{\mu}_m})}D_{n_{\bm{\nu}_m}}^{(l_{\bm{\nu}_m})}
  \left[(2l_{\bm{\mu}_m}+1)\frac{(l_{\bm{\mu}_m}-m)!}
  {(l_{\bm{\mu}_m}+m)!}\right]^{1/2}
  \left[(2l_{\bm{\nu}_m}+1)\frac{(l_{\bm{\nu}_m}-m)!}
  {(l_{\bm{\nu}_m}+m)!}\right]^{1/2}\ .
\end{equation}

Gradient potentials  $\hat{V}^{(r)}$ and $\hat{V}^{(\vartheta)}$ are more suited for calculation of their matrix elements in the spherical coordinate frame located at the center of the inner sphere. This is due to the fact that in the definitions of these potentials, Eqs.~\eqref{V^(r)-def} and \eqref{V^(theta)-def}, there are present derivatives of function $\varepsilon(r,\vartheta)$ with respect to both radial and polar coordinates. These derivatives make the potentials $\hat{V}^{(r)}$ and $\hat{V}^{(\vartheta)}$ singular, namely, proportional to $\delta$-functions arisen when crossing the boundary of the inner sphere. Although singularities of this form are not problematic for matrix element calculations, to concretize them for a specific position of the inner sphere it is advantageous to go to the reference system related just to this sphere center. For this purpose we can make use of the gage preserving coordinate transformation
\begin{equation}\label{ChangeVar}
\begin{split}
  & r\sin\vartheta = r'\sin\vartheta'\ ,\\
  & r\cos\vartheta = r'\cos\vartheta'+d\ ,
\end{split}
\end{equation}
where the primes relate to the inner-sphere coordinate frame. In this system, the logarithmic derivatives entering into Eqs.~\eqref{V^(r)-def} and \eqref{V^(theta)-def} are calculated to
\begin{subequations}\label{ln_derivs}
\begin{align}
\label{d_ln_e/dr}
 & \left[\frac{\partial\ln\varepsilon(r,\vartheta)}{\partial r}\right](\vartheta',r')=
  2\,\frac{\varepsilon_{\mathrm{out}}-\varepsilon_{\mathrm{in}}}
  {\varepsilon_{\mathrm{out}}+\varepsilon_{\mathrm{in}}}\cdot
  \frac{R_{\mathrm{in}}+d\cos\vartheta'}
  {\sqrt{R_{\mathrm{in}}^2+2R_{\mathrm{in}}d\cos\vartheta'+d^2}}
  \delta(r'-R_{\mathrm{in}})\ ,
  \\
\label{d_lne/d_theta}
 & \left[\frac{\partial\ln\varepsilon(r,\vartheta)}{\partial\vartheta}\right](\vartheta',r')=
  2\,\frac{\varepsilon_{\mathrm{out}}-\varepsilon_{\mathrm{in}}}
  {\varepsilon_{\mathrm{out}}+\varepsilon_{\mathrm{in}}}\,
  d\,\sin\vartheta'\delta(r'-R_{\mathrm{in}})\ .
\end{align}
\end{subequations}
When deriving the latter two formulas, we have put the permittivity value on the border between regions with $\varepsilon_{\mathrm{in}}$ and $\varepsilon_{\mathrm{out}}$ equal to their half-sum.

With the use of \eqref{ln_derivs}, mode matrix elements of both of the gradient potentials entering into Eq.~\eqref{u_0-eq} are calculated to
\begin{subequations}\label{Grad_pots}
\begin{align}
\label{Grad_(r)}
  \mathcal{U}_{\bm{\mu}_0\bm{\nu}_0}^{(r)}=&
  \frac{B_{\bm{\mu}_0\bm{\nu}_0}}{R_{\mathrm{out}}^2}\cdot
  \frac{\varepsilon_{\mathrm{out}}-\varepsilon_{\mathrm{in}}}
  {\varepsilon_{\mathrm{out}}+\varepsilon_{\mathrm{in}}} I_{\bm{\mu}_0\bm{\nu}_0}^{(r)}\ ,\\
\label{Grad_(theta)}
  \mathcal{U}_{\bm{\mu}_0\bm{\nu}_0}^{(\vartheta)}=&
  2M_{\mathrm{in}}\frac{B_{\bm{\mu}_0\bm{\nu}_0}}{R_{\mathrm{out}}^2}\cdot
  \frac{\varepsilon_{\mathrm{out}}-\varepsilon_{\mathrm{in}}}
  {\varepsilon_{\mathrm{out}}+\varepsilon_{\mathrm{in}}} I_{\bm{\mu}_0\bm{\nu}_0}^{(\vartheta)}\ ,
\end{align}
\end{subequations}
where the integrals $I_{\bm{\mu}_0\bm{\nu}_0}^{(r)}$ and $I_{\bm{\mu}_0\bm{\nu}_0}^{(\vartheta)}$ have the form
\begin{subequations}\label{Integrals(r)(theta)}
\begin{align}
\label{Integral(r)}
  I_{\bm{\mu}_0\bm{\nu}_0}^{(r)} &= \int\limits_{-1}^1 dt\,
  \frac{1+M_{\mathrm{in}}t}{\xi^2\big(t;M_{\mathrm{in}}\big)}
  P_{l_{\bm{\mu}_0}}^{0}\bigg[\frac{t+M_{\mathrm{in}}}
  {\xi\big(t;M_{\mathrm{in}}\big)}\bigg]
  P_{l_{\bm{\nu}_0}}^{0}\bigg[\frac{t+M_{\mathrm{in}}}
  {\xi\big(t;M_{\mathrm{in}}\big)}\bigg]
  J_{l_{\bm{\mu}_0}+\frac{1}{2}}
  \left(\lambda_{n_{\bm{\mu}_0}}^{(l_{\bm{\mu}_0})}\frac{R_{\mathrm{in}}}{R_{\mathrm{out}}}
  \xi\big(t;M_{\mathrm{in}}\big)\right)
  \notag\\
  &\quad\times  \Bigg\{\frac{1}{\xi\big(t;M_{\mathrm{in}}\big)}
  J_{l_{\bm{\nu}_0}+\frac{1}{2}}
  \left(\lambda_{n_{\bm{\nu}_0}}^{(l_{\bm{\nu}_0})}\frac{R_{\mathrm{in}}}{R_{\mathrm{out}}}
  \xi\big(t;M_{\mathrm{in}}\big)\right)+
  \notag\\
  & \qquad\quad
  \phantom{\frac{1}{\xi\big(t;M_{\mathrm{in}}\big)}}
  + \lambda_{n_{\bm{\nu}_0}}^{(l_{\bm{\nu}_0})}\frac{R_{\mathrm{in}}}{R_{\mathrm{out}}}
  \bigg[J_{l_{\bm{\nu}_0}-\frac{1}{2}}
  \left(\lambda_{n_{\bm{\nu}_0}}^{(l_{\bm{\nu}_0})}\frac{R_{\mathrm{in}}}{R_{\mathrm{out}}}
  \xi\big(t;M_{\mathrm{in}}\big)\right)-
  \notag\\
  & \qquad\quad
  \phantom{\frac{1}{\xi\big(t;M_{\mathrm{in}}\big)}}
  \phantom{\lambda_{n_{\bm{\nu}_0}}^{(l_{\bm{\nu}_0})}\frac{R_{\mathrm{in}}}{R_{\mathrm{out}}} \bigg[J-}
  - J_{l_{\bm{\nu}_0}+\frac{3}{2}}
  \left(\lambda_{n_{\bm{\nu}_0}}^{(l_{\bm{\nu}_0})}\frac{R_{\mathrm{in}}}{R_{\mathrm{out}}}
  \xi\big(t;M_{\mathrm{in}}\big)\right)\bigg]\Bigg\}\ ,\\
\label{Integral(r)}
  I_{\bm{\mu}_0\bm{\nu}_0}^{(\vartheta)} &=
  \int\limits_{-1}^{1}
  \frac{dt}{\Big[\xi\big(t;M_{\mathrm{in}}\big)\Big]^{3}}
  J_{l_{\bm{\mu}_0}+\frac{1}{2}}
  \left[\lambda_{n_{\bm{\mu}_0}}^{(l_{\bm{\mu}_0})}\frac{R_{\mathrm{in}}}{R_{\mathrm{out}}}
  \xi\big(t;M_{\mathrm{in}}\big)\right]
  J_{l_{\bm{\nu}_0}+\frac{1}{2}}
  \left[\lambda_{n_{\bm{\nu}_0}}^{(l_{\bm{\nu}_0})}\frac{R_{\mathrm{in}}}{R_{\mathrm{out}}}
  \xi\big(t;M_{\mathrm{in}}\big)\right] \times
\notag\\
 & \quad\times P_{l_{\bm{\mu}_0}}^0\bigg[\frac{t+M_{\mathrm{in}}}{\xi\big(t;M_{\mathrm{in}}\big)}\bigg]
 \Bigg\{
  P_{l_{\bm{\nu}_0}}^0\bigg[\frac{t+M_{\mathrm{in}}}{\xi\big(t;M_{\mathrm{in}}\big)}\bigg]
  \big(t+M_{\mathrm{in}}\big)-
\notag\\
  & \quad
  \phantom{P_{l_{\bm{\mu}_0}}^0\bigg[\frac{t+M_{\mathrm{in}}}{\xi\big(t;M_{\mathrm{in}}\big)}\bigg]
 \Bigg\{ P_{l_{\bm{\nu}_0}}^0\bigg[\frac{t+M_{\mathrm{in}}}{\xi\big(t;M_{\mathrm{in}}\big)}\bigg]}
  -\left(P_{l_{\bm{\nu}_0}}^0\right)'\bigg[\frac{t+M_{\mathrm{in}}}
  {\xi\big(t;M_{\mathrm{in}}\big)}\bigg]
  \frac{1-t^2}{\xi\big(t;M_{\mathrm{in}}\big)}\Bigg\}\ .
\end{align}
\end{subequations}
In Eqs.~\eqref{Integrals(r)(theta)}, function $\xi\big(t;M_{\mathrm{in}}\big)= \sqrt{1+2tM_{\mathrm{in}}+M_{\mathrm{in}}^2}$ is the scaling factor that represents the dimensionless (in units of $R_{\mathrm{in}}$) length of radius-vector $\mathbf{r}\in\Sigma_{\mathrm{out}}$, ${M_{\mathrm{in}}=d/R_{\mathrm{in}}}$ is the dimensionless (in the same units) misalignment of the dielectric spheres of the resonator infill.
}

%%%%%%%%%%%%%%%%%%%%%%%%%%%%%%%%%%%%%%%
\bibliographystyle{Vancouver}
\bibliography{scrrefs}
%%%%%%%%%%%%%%%%%%%%%%%%%%%%%%%%%%%%%%%
 \end{document}